\documentclass[reprint,superscriptaddress,aps,prl]{revtex4-2}
\usepackage[T1]{fontenc}
\usepackage[latin1]{inputenc}
\usepackage{lmodern}
\expandafter\let\csname equation*\endcsname\relax
\expandafter\let\csname endequation*\endcsname\relax
\usepackage{amsmath,physics}
\usepackage{amssymb,dsfont}
\usepackage{graphicx}
\usepackage{xcolor}
\usepackage{natbib}
\usepackage{bm}
\usepackage{hyperref}
\usepackage{bbold}
\usepackage[mathscr]{euscript}
\usepackage[cal=boondoxo]{mathalfa}
\usepackage{comment}
\renewcommand\footnotemark{}

\usepackage{braket}
\usepackage[normalem]{ulem}
\usepackage{lineno}

\def\ii{{\rm i}} 

\begin{document}

\title{Emergent non-Markovianity in time-delayed waveguide QED}
\author{Debsuvra Mukhopadhyay}
\email{debsuvram@usf.edu}
\author{Stuart J. Masson}
\email{sjmasson@usf.edu}
\affiliation{Department of Physics, University of South Florida, Tampa, Florida 33620, USA}
\date{\today}

\begin{abstract}
Quantum systems are invariably coupled to a surrounding environment. A common theoretical method to simplify the problem is to trace over the environment under a Markov approximation, which assumes that the environment holds no memory over the timescales that the system evolves under. In interacting many-body systems, it is often not trivial to extract which of the multiple collective timescales are relevant. We consider a time-delayed waveguide QED setup with a single excitation. By fixing the maximum propagation time and increasing the number of emitters, we show that, even for small delay times, collective effects are sufficient to cause non-Markovianity. The impact of memory is intrinsically state-dependent. For superradiant states, the relevant system timescale is the superradiant lifetime while the relevant bath timescale is the end-to-end delay time. For subradiant states, both timescales depend on the structure of the specific state. Our results demonstrate the importance of prudently making Markov approximations in quantum many-body systems, and highlight potential pitfalls to avoid in scaling up quantum devices to large system sizes.
\end{abstract}

\maketitle

Open quantum systems are ubiquitous because no quantum system can be perfectly isolated from its environment~\cite{BreuerBook,RivasBook}. Understanding dissipation is therefore central to quantum science: environmental coupling degrades coherence and limits lifetimes, but can also serve as a resource for mediating interactions, stabilizing states, and transferring quantum information between distant nodes~\cite{Poyatos96,Cirac97,Verstraete09}. Open quantum systems are most commonly studied under a Markov approximation, where the environment is assumed to have no memory. The Markov approximation allows the environment to be traced over replacing its effects with time-local coherent interactions and dissipation~\cite{Carmichael93,Stefanini26}. Determining when the Markov approximation is appropriate and when environmental memory must be explicitly retained is essential for predictive quantum simulation, control, and dissipation engineering. For a single emitter coupled to an environment, the criterion is intuitive: the environment memory time must be very short compared with the radiative lifetime, such that the emitted photon is effectively instantaneously lost.

\begin{figure}[b!]
    \centering
    \includegraphics[width=\columnwidth]{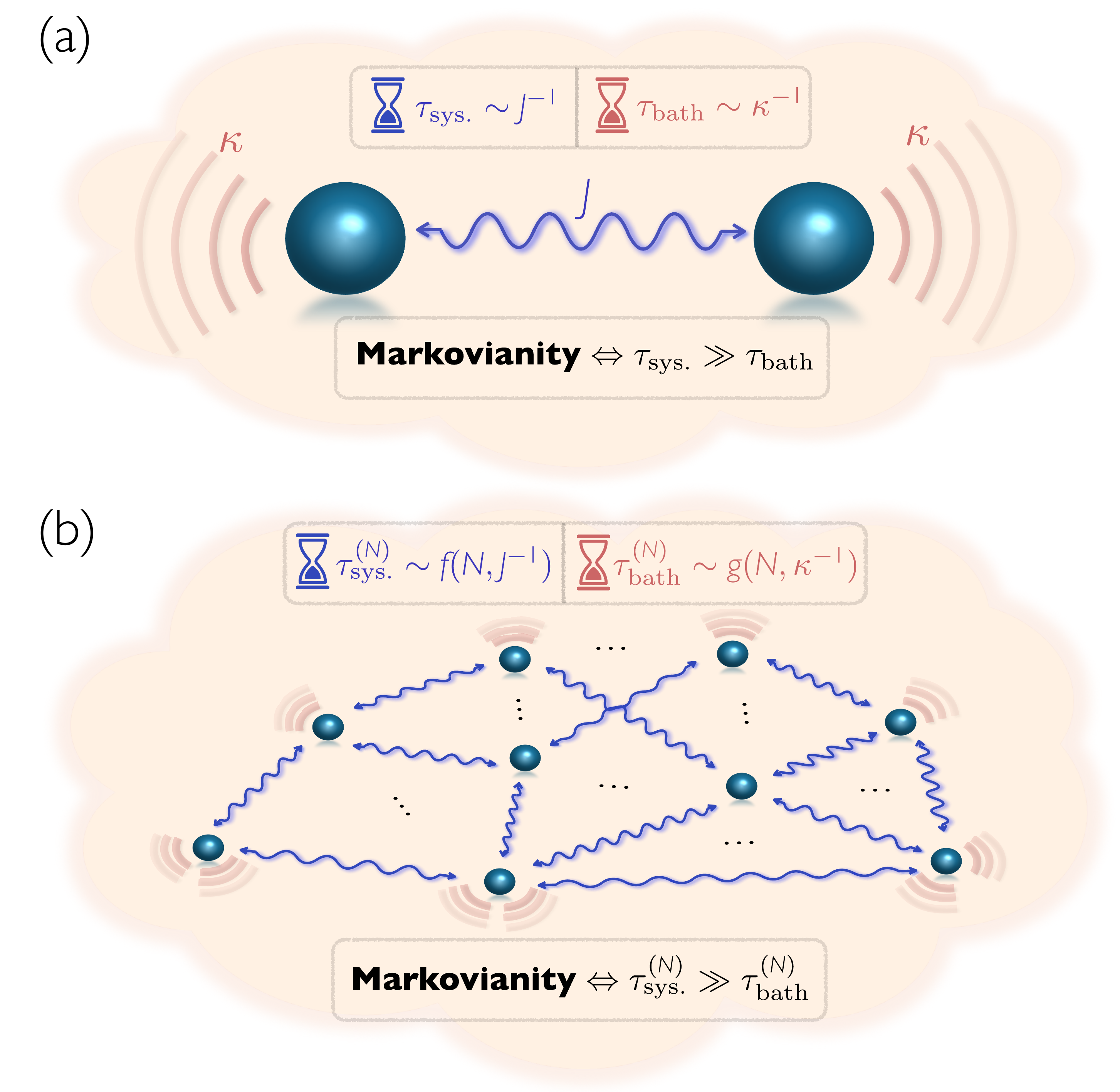}
    \caption{Markov approximation in many-body systems. (a)~In bipartite systems, the relevant timescales for the Markov approximation to be valid are readily identified. The relevant system timescale is determined by the interaction rate between the particles, $J$, and the relevant environmental timescale by how fast bath correlations are erased, $\kappa$. (b) For $N$ particles, the relevant timescales are less obvious. Both timescales are still determined by the microscopic rates but can also be modified by system size.
    }
    \label{fig:intro_figure}
\end{figure}

Identifying the relevant environment and system timescales in many-body open quantum systems is far less straightforward, as illustrated in Fig.~\ref{fig:intro_figure}. Many-body light--matter interfaces are governed by collective modes whose radiative lifetimes can differ greatly from those of their individual constituents~\cite{Dicke54}. At the same time, spatially extended environments introduce a hierarchy of memory times between different points in the system. Emergent effects arise across diverse quantum platforms, from superradiant bursts~\cite{Gross82,BenedictBook,Masson22}, to long-lived states stabilized by interference~\cite{Palma96,Ramos14,Asenjo17PRX}, and driven-dissipative phase transitions~\cite{Baumann10,Fitzpatrick17,Benary22}. As quantum technologies increase in complexity, it becomes increasingly important to understand the role of many-body dynamics in the timescales underpinning the Markov approximation. 

Environmental memory due to propagation delay reduces cooperation in collective radiative decay. Early work introduced the maximum cooperation number as a collective organizing scale~\cite{Arecchi70,Bonifacio75}. Only emitters which fall within the propagating light cone over the lifetime of the state will behave collectively, leading to deviations from fully-collective behavior when the product of the maximum propagation time and collective radiative rate per atom becomes comparable to unity~\cite{Bonifacio75}. While the predicted scaling was numerically identified for many-body superradiance into a chiral waveguide~\cite{Windt25}, our results highlight that following this logic for arbitrary states is incorrect.

 Waveguide quantum electrodynamics (wQED) provides a natural setting for studying memory effects arising from finite photon propagation~\cite{Zheng13,GonzalezBallestero13}. For superradiant states, retardation can enhance collective decay faster than Dicke superradiance~\cite{Dinc19PRR,Dinc19Quantum,Sinha20,BarahonaPascual25,Capurso26} and lead to anomalous population of dark states~\cite{Calajo19,Carmele20,AlvarezGiron24,Magnifico25,Capurso26}. For subradiant states, retardation causes transient leakage~\cite{Sinha20,Guha25} and changes the structure~\cite{Vera26arxiv}. Such phenomena are especially prominent in slow-propagation reservoirs, including waveguides near photonic band edges~\cite{Mirhosseini18,Scigliuzzo22,Goban15,Tiranov23}, surface-acoustic-wave transmission lines~\cite{Gustafsson14,Dumur21} and atomic-matter-wave baths~\cite{Krinner18,Kim25}, as well as for short-lived emitters~\cite{Lodahl15,Wang18,Feldmann87}. These works establish the interplay between collective effects and non-Markovian behavior, motivating the need for a general criterion to assess the validity of the Markov approximation with increasing system size.

\begin{figure}[t]
    \centering
    \includegraphics[width=\columnwidth]{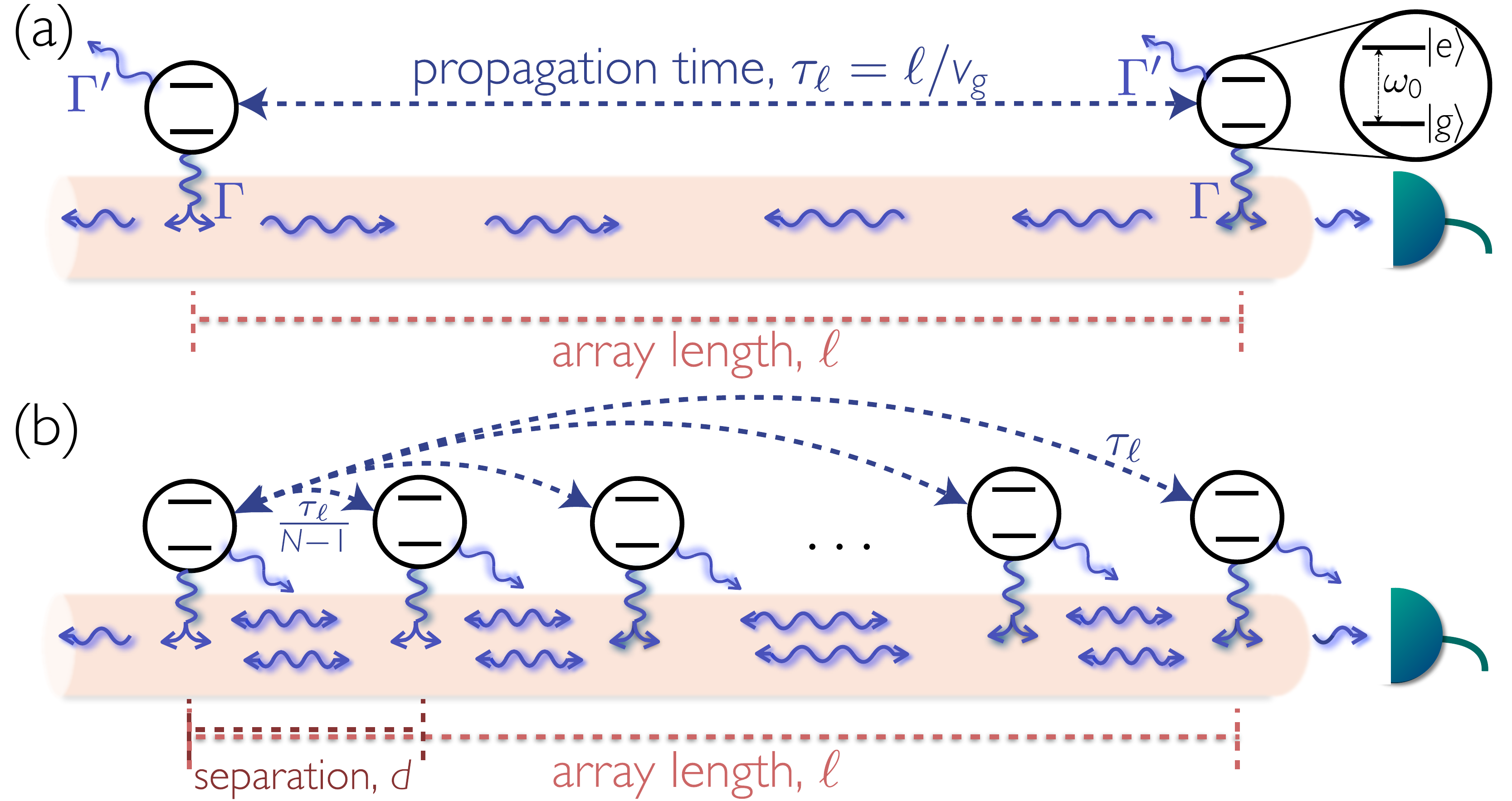}
    \caption{Setup. (a) two to (b) $N$ emitters coupled to a waveguide spaced evenly at $k_0d = 2\pi q$ with total length $\ell$ and thus total propagation time $\tau_l = \ell / v_g$, where $v_g$ is the group velocity. Each emitter is coupled to the waveguide at rate $\Gamma$ and to free space at rate $\Gamma'$. 
    }
    \label{fig:schematic_array}
\end{figure}

 In this manuscript, we establish scaling laws for the breakdown of Markovianity in time-delayed single-excitation waveguide QED. We find that collective effects activate non-Markovian dynamics by varying the emitter number at fixed array length. Memory is intrinsically state-selective: the same delay may be negligible for one state yet decisive for another. In the bright sector, the ratio of the fixed traversal time to the Markovian superradiant lifetime dictates dynamics. In the dark sector, the impact of retardation depends on the structure of the dark states; some states obey the same scaling as superradiant states with increasing emitter number while others become more similar to their Markovian counterparts. Emitter number and collective-state structure become control parameters for time-delayed collective dynamics. More broadly, our results highlight that care must be taken to identify the relevant emergent timescales when making a Markov approximation in a many-body system.

\textit{System.---} We consider $N$ two-level emitters of transition frequency $\omega_0$ coupled to a bidirectional, one-dimensional waveguide, as illustrated in Fig.~\ref{fig:schematic_array}. Within the collective-emission bandwidth, the guided modes are taken as linearly dispersive with group velocity $v_g$. Under the rotating-wave approximation, the interaction-picture Hamiltonian is given by $(\hbar=1)$
\begin{equation}
\begin{aligned}
    \hat H_{\rm int}(t)
    &=
    \sum_{m=1}^{N}\sum_{\mu=\pm}
    \int_0^\infty d\omega\,
    \Big[
        g(\omega)\hat{\sigma}_+^{(m)}
        \hat{a}_{\mu,\omega}  \\
    &\hspace{2cm}\times
        e^{-\ii(\omega-\omega_0)t}
        e^{\ii\mu\omega x_m/v_g}
        +{\rm H.c.}
    \Big].
\end{aligned}
\label{eq:interaction_picture_hamiltonian}
\end{equation}
Here $\hat a_{\mu,\omega}$ annihilates a guided photon of frequency $\omega$ in the propagation direction $\mu$, with $\mu=+(-)$ labeling the right-(left-) moving mode, and $\hat\sigma_\pm^{(m)}$ denote the raising and lowering operators of the emitter at position $x_m$.

The Hamiltonian in Eq.~\eqref{eq:interaction_picture_hamiltonian} conserves excitation number, such that the preparation of a single collective excitation, $\ket{\psi_{\rm at}(0)}=\sum_{m=1}^Nc_m(0)\hat{\sigma}_+^{(m)} \ket{g}^{\otimes N}$ and the waveguide in vacuum, leads to evolution entirely within the single-excitation sector. The dynamics is then fully described by the emitter amplitudes and the one-photon wavefunction. Eliminating the guided continuum using a slowly varying coupling $g(\omega)$, while retaining the finite photon flight times, shows that the field emitted by emitter $n$ reaches emitter $m$ after a delay $\tau_{mn}=|x_m-x_n|/v_g$, and acquires the phase $e^{\ii k_0|x_m-x_n|}$, with $k_0=\omega_0/v_g$. We choose an equally spaced chain, $x_m=x_1+(m-1)d$, of length $\ell=(N-1)d$ in the phase-matched geometry $k_0d=2\pi q$, with $q\in\mathbb Z$, to remove the static propagation phases. Neglecting nonguided losses, the emitter amplitudes then obey~\cite{Sinha20}
\begin{equation}
\dot c_m(t)
=
-\frac{\Gamma}{2}c_m(t)
-
\frac{\Gamma}{2}
\sum_{n\neq m}
c_n(t-\tau_{mn})
\Theta(t-\tau_{mn}),
\label{eq:delay_eom_inphase}
\end{equation}
where $\Gamma$ is the single-emitter coupling to the waveguide. The Heaviside function $\Theta(t-\tau_{mn})$ encodes the causal structure. Equation~\eqref{eq:delay_eom_inphase} is the minimal dynamical model used throughout the main text, with generalizations addressed in the Supplemental Material~\cite{Si_enm}.

Applying the Born--Markov approximation directly to Eq.~\eqref{eq:interaction_picture_hamiltonian} yields, in the same phase-matched geometry, the master equation $\dot{\hat{\rho}}=\Gamma\,(\hat S_-\hat\rho\hat S_+ -\{\hat S_+\hat S_-,\hat\rho\})/2$~\cite{Chang12}, with $\hat S_\pm=\sum_{m=1}^{N}\hat\sigma_\pm^{(m)}$ and $\hat{\rho}$ the emitter density matrix. The symmetric state $\ket B=N^{-1/2}\hat{S}_+\ket{g}^{\otimes N}$ is bright and exponentially decays at a rate $N\Gamma$, while the $N-1$ orthogonal states form a degenerate perfectly dark manifold. We probe deviations from these Markovian dynamics as a function of emitter number while holding the chain length $\ell$ fixed, thereby isolating the role of collective scaling.

\begin{figure}[t]
\centering
\includegraphics[width=\columnwidth]{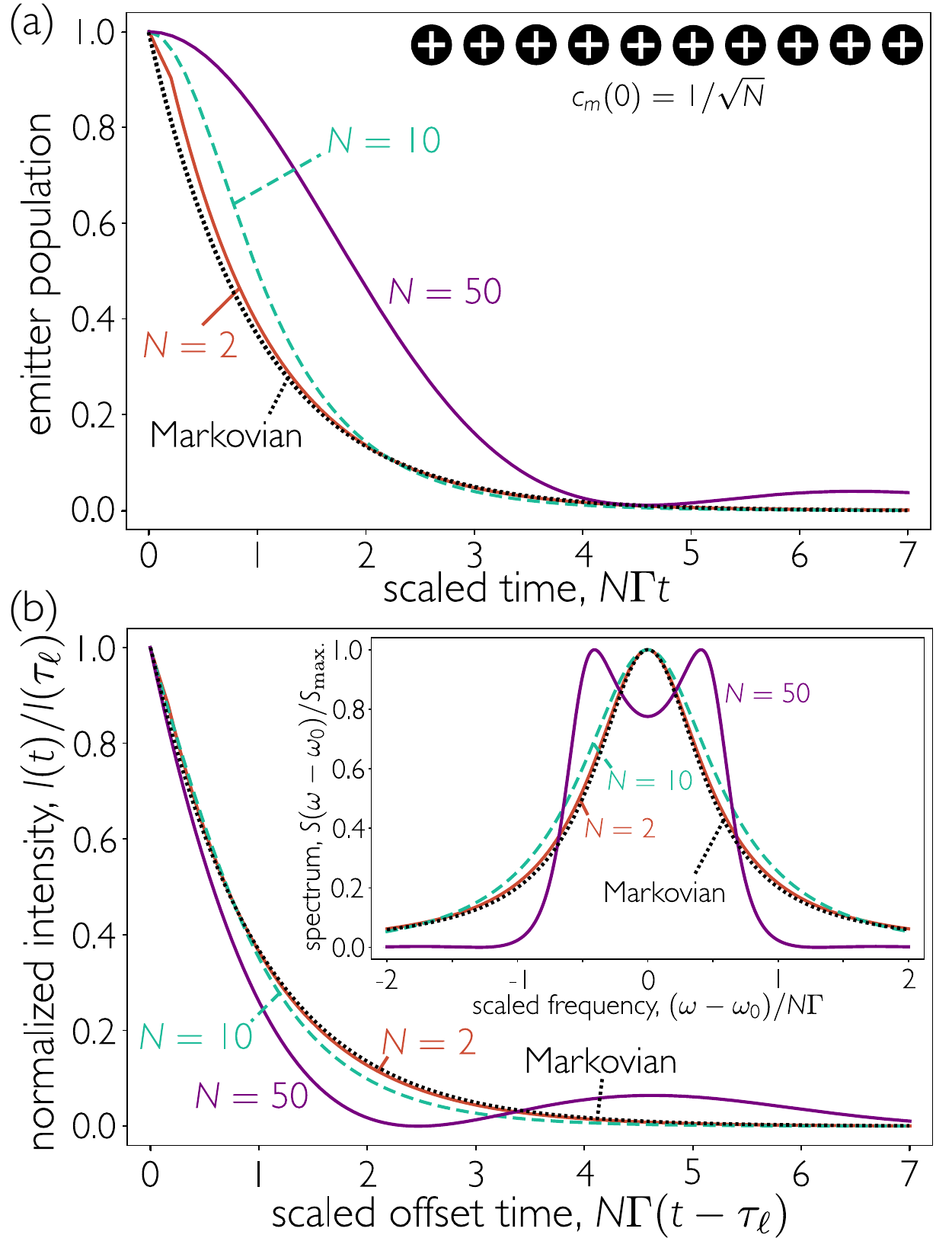}
\caption{
Emergent non-Markovian dynamics for superradiant states. (a) Time evolution of total emitter population, $\sum_{m=1}^N |c_m(t)|^2$, for different emitter numbers. (b) Intensity registered by a photon detector at the position of the right-most emitter, offset by the delay time, for different emitter numbers. (inset) Power spectrum of the emitted light, normalized by the maximum spectral weight, for different emitter numbers. In all plots, the dotted black line represents the $N$-independent Markovian limit, and simulations use a total end-to-end delay of $\Gamma\ell / v_g = 0.1$ and an initial single-excitation superradiant state $\ket{\psi(t=0)} =  \hat{S}_+ \ket{g}^{\otimes N} / \sqrt{N}$.}
\label{fig:bright_memory_signatures}
\end{figure}

\textit{Results.---} Retardation has a qualitatively different consequence for different initial states: the relevant timescales depend on the state structure. The bright sector exhibits a breakdown of exponential decay as the emitter number is increased. Fig.~\ref{fig:bright_memory_signatures}(a) shows that for very small arrays, the evolution remains essentially Markovian. For moderate values of $N$, the population monotonically decays but displays noticeable deviations from the Markov reference. For sufficiently large arrays, retardation qualitatively reshapes the dynamics inducing population revivals. These revivals are a direct consequence of environmental memory: radiation previously emitted into the waveguide remains accessible to the array and is partially reabsorbed. The same emergent non-Markovian effects are also present with arbitrary emitter spacings, i.e., $k_0d \neq 2\pi q$~\cite{Si_enm}.

 Memory effects also manifest in the emitted field, as shown in Fig.~\ref{fig:bright_memory_signatures}(b). In the Markovian limit, the bright state emits an exponentially decaying wave packet with a Lorentzian spectrum of full width $N\Gamma$. Small arrays remain close to this reference, while an intermediate $N$ exhibits only modest temporal and spectral deviations. At large $N$, however, the response is qualitatively different: the output intensity develops a delayed revival, while the spectrum splits into two resolved peaks [see Supplemental Material for definitions~\cite{Si_enm}]. The splitting is not due to vacuum-Rabi splitting, but instead, finite propagation causes the emitter-containing waveguide segment to behave as an effective multimode resonator, with characteristic resonance spacing $\tau_{\ell}^{-1}=v_g/\ell$~\cite{Sinha20PRA,Cilluffo26,Dhara26arxiv} reminiscent of emitter-defined cavities in wQED~\cite{Chang12,Mirhosseini19}. Because $\ell$ is fixed, this spacing remains constant, while the Markovian bright-state linewidth $N\Gamma$ grows linearly with $N$ until it resolves spectral peaks, providing a direct spectral signature of memory~\cite{Sinha20PRA,Keefe25}. 

\begin{figure}[t]
    \centering
    \includegraphics[width=\columnwidth]{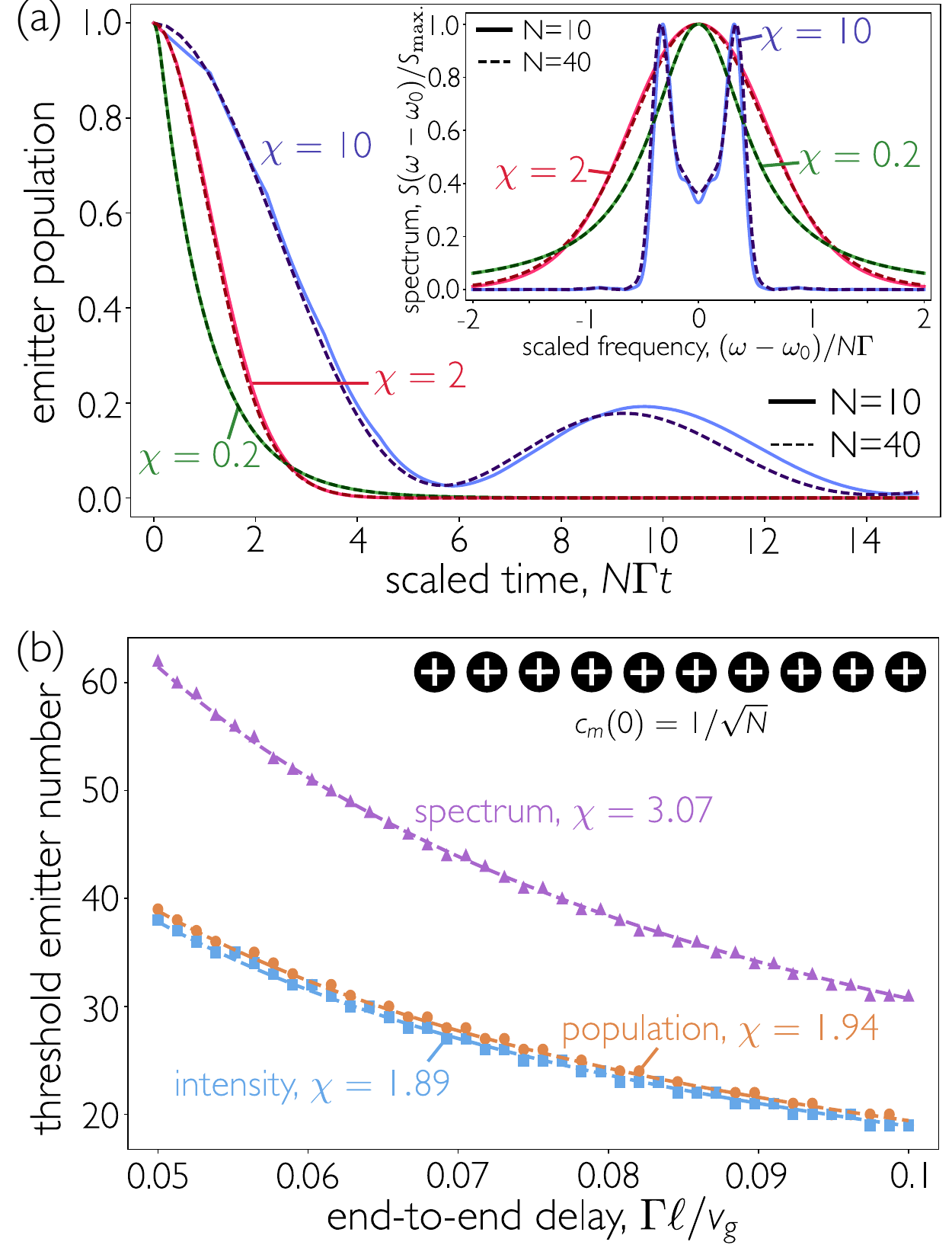}
    \caption{
    Extracting relevant timescales for Markovianity. (a) Time evolution of total emitter population, $\sum_{m=1}^N |c_m(t)|^2$, for different values of $\chi = N\Gamma\ell/v_g$ with $N=10$ (solid lines) and $N=40$ (dashed).
    (inset) Power spectrum of the emitted light, normalized by the maximum spectral weight, for different values of $\chi$ with $N=10$ (solid lines) and $N=40$ (dashed). (b) Minimum emitter number for which certain features appear at fixed end-to-end delay. These features are two peaks in the spectrum (purple triangles), revivals in the emitted intensity (blue squares), and revivals in the emitter population (orange circles). All datasets are fitted to lines of constant $\chi$ (dashed lines). Both plots use an initial single-excitation superradiant state $\ket{\psi(t=0)} = \hat{S}_+ \ket{g}^{\otimes N} / \sqrt{N}$.
    }
    \label{fig:chi_thresholds}
\end{figure}

Non-Markovian dynamics of the superradiant state at arbitrary delays are solely controlled by an emergent parameter $\chi\equiv {N\Gamma \ell}/{v_g} = N\Gamma\tau_{\ell}$ in the large $N$ limit, as shown in Fig.~\ref{fig:chi_thresholds}(a). For $\chi=N\Gamma\ell/v_g\lesssim \mathcal{O}(1)$, time traces corresponding to different emitter numbers and propagation delays nearly collapse when time is measured in units of the superradiant lifetime $\tau_B = (N\Gamma)^{-1}$. The corresponding spectra, shown in the inset, exhibit an analogous collapse. Together, these observations demonstrate that the time-domain and spectral responses are both governed primarily by the product $N\Gamma\ell/v_g$, rather than by $N$ or $\ell$ separately. Note that if the spacing between emitters, $d$, is kept constant as the system size is increased, then the dynamics are controlled by $\chi \approx N^2\Gamma d / v_g$. At larger $\chi$, finite-size effects play a more significant role but the asymptotic collapse remains. This universal scaling holds for imperfect and chiral couplings, as shown in the Supplemental Material~\cite{Si_enm}.

The structure of the memory parameter $\chi=\tau_{\ell}/\tau_B$ identifies both the memory and system timescales that govern the crossover. The relevant memory timescale is the end-to-end traversal time $\tau_{\ell}$, not the nearest-neighbor delay, which decreases as the array is made denser at fixed $\ell$; likewise, the relevant system timescale is the collective lifetime $\tau_B$, rather than the bare single-emitter lifetime. 

The relevant timescales are not those obtained by a description of cooperative emission based on the decay rate per emitter~\cite{Arecchi70,Bonifacio75}. For the single-excitation superradiant state, the decay rate per emitter is constant, yet behavior changes significantly with $N$. Instead, the relevant timescale can be obtained either by considering the temporal width of the emitted pulse or the fastest decay rate in the system, both of which scale as $N\Gamma$. For many-body superradiance, the temporal width scales with $N\Gamma$ while the fastest decay rate scales as $N^2\Gamma$. The $N\Gamma$ scaling found for a chiral waveguide~\cite{Windt25} should not be explained by the decay rate per emitter but taken as an indication that the relevant timescale is the temporal width of the emitted pulse.

\begin{figure*}[t]
\centering
\includegraphics[width=\textwidth]{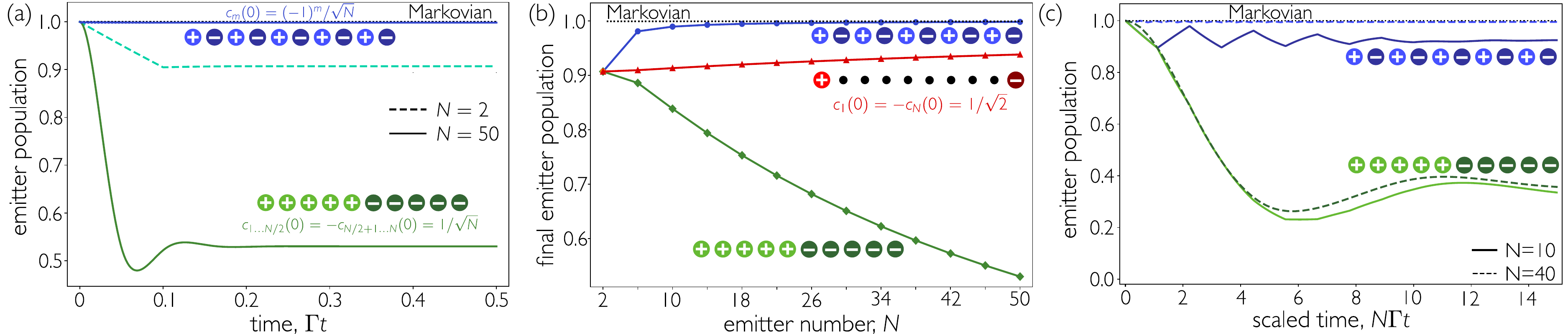}
\caption{
Emergent non-Markovian dynamics for subradiant states. (a) Time evolution of total emitter population, $\sum_{m=1}^N |c_m(t)|^2$, for different subradiant states. The initial state has equal amplitude on all emitters with (green) two domains i.e., $\ket{\psi(t=0)} = (\sum_{m=1}^{N/2} \hat{\sigma}_+^{(m)} - \sum_{m=N/2+1}^N \hat{\sigma}_+^{m}) \ket{g}^{\otimes N}/ \sqrt{N}$ and (blue) alternating signs i.e., $\ket{\psi(t=0)} = \sum_{m=1}^N (-1)^{(m)} \hat{\sigma}_+^{(m)} \ket{g}^{\otimes N} / \sqrt{N}$ with $N=50$. Both states have the same form for $N=2$, shown in turquoise. (b) Scaling of the equilibrium dark state population with emitter number for three different subradiant states: two domains (green diamonds), alternating signs (blue circles), and a subradiant superposition of the two end emitters, i.e., $\ket{\psi(t=0)} = (\hat{\sigma}_+^{(1)} - \hat{\sigma}_+^{(N)}) \ket{g}^{\otimes N} / \sqrt{2}$. (c) Time evolution of total emitter population for different subradiant states with fixed $\chi=10$ and $N=10$ (solid lines) and $N=40$ (dashed). In (a,b), the total end-to-end delay is $\Gamma\ell / v_g = 0.1$. In all plots, the dotted black line represents the $N$-independent Markovian limit.
}
\label{fig:subradiant}
\end{figure*}

The validity of the Markov approximation in single-excitation superradiance requires $\chi \ll 1$. For small delay and large arrays, $N\gg1$, expanding the superradiant decay rate at fixed chain length gives [see~\cite{Si_enm} for details]
\begin{equation}
\Gamma_{B}(\chi)
\simeq
N\Gamma
\left(
1+\frac{\chi}{6}
+\frac{\chi^2}{20}
+\mathcal{O}(\chi^3)
\right).
\label{eq:SR_second_order_fixed_length}
\end{equation}
The memory parameter $\chi$ also controls the onset of strongly non-Markovian behavior. By fixing the end-to-end delay and increasing emitter number, we find a threshold number of emitters at which revivals appear in the emitter populations and field intensity, and the spectrum splits into two distinct peaks, as shown in Fig.~\ref{fig:chi_thresholds}(b). All three thresholds converge on the same scaling: they occur at constant $\chi$.

The validity of the Markov approximation in the dark sector depends on the particular state. Destructive interference cannot be established instantaneously in the presence of delay and instead takes effect when propagating light interferes with light of the opposite phase, as shown in Fig.~\ref{fig:subradiant}(a). An initial state with equal amplitudes on each emitter but alternating signs, i.e. $c_m(0)\propto(-1)^{m-1}$, is subradiant at the level of nearest neighbors. Destructive interference is established at a time $\sim1/N$ before which emitters behave independently. This sets a constant system timescale and a decreasing environmental timescale, such that the alternating-sign states behave closer to the Markovian limit with increasing system size. The double-domain family of states, where the left and right halves of the array have opposite signs, retains the same two-emitter limit, but behaves in the opposite manner. Destructive interference is established after radiation has propagated between the domains, meaning that the relevant environmental timescale is unchanged with increasing system size. At intermediate times, each block constructively interferes leading to superradiant emission and progressively greater leakage with $N$ before inter-domain interference becomes effective so the relevant system timescale decreases. This results in fundamentally different scaling with $N$ for different states, as shown in Fig.~\ref{fig:subradiant}(b). Another example is the edge-localized antisymmetric state. Here, destructive interference is established over a timescale corresponding to the whole array but without superradiant enhancement, such that behavior shows weak dependence on $N$.

Subradiant states with excitation patterns that can be consistently mapped to a well-defined density function as $N\rightarrow\infty$ are also controlled by $\chi$ in the large-$N$ limit and thus must behave less like their Markovian counterparts when $N$ is increased at fixed $\ell$. For example, the double-domain state is controlled by $\chi$ as it is well-defined by a piecewise flat density function that switches sign halfway along the array. Any well-defined density function must at large-enough $N$ feature blocks of constructively interfering emitters and thus exhibit the same asymptotic scaling. States which do not have such a mapping can exhibit the opposite behavior, as shown in Fig.~\ref{fig:subradiant}(c).

\textit{Conclusion---} Our findings offer a cautionary lesson for many-body open quantum systems: the validity of the Markov approximation is not fixed by the environment alone, but can change qualitatively with system size, collective-state structure, and which states are dynamically populated. In the bright sector, collective enhancement shortens the relevant system timescale until finite propagation cannot be neglected, with the crossover organized by the ratio between the end-to-end propagation time and the superradiant lifetime: $\chi=\tau_{\ell}/\tau_B$. The situation is subtler in the dark sector, where some states are organized by $\chi$ and others are not. Thus, even within a nominally dark manifold, the validity of a Markov treatment depends on the collective state used to encode quantum information. Our findings provide access to large-$N$ behavior of single-excitation states in time-delayed waveguide QED: if the initial state can be suitably generalized to fewer emitters then rescaled dynamics can be calculated for the same $\chi$.

Emergent non-Markovianity has direct implications for quantum technologies based on collective effects. Superradiant performance gains~\cite{Paulisch19,Koppenhofer22,Bohr24} cannot be extrapolated indefinitely with emitter number: retardation can initially accelerate collective emission, but stronger memory effects cause reabsorption, revivals, and multimode emission. Our results also mean that technologies using single dark states~\cite{Asenjo17PRX,CastellsGraells25,Zafra26} should focus on certain states with favorable scaling. However, technologies that require multiple dark states, such as decoherence-free subspaces~\cite{Zanardi97,Lidar98}, would require all those states and their superpositions to scale favorably. Postselection via photodetection to erase leakage errors fails because such errors are biased toward particular collective states and so backaction in selected runs would alter the state.

The single-excitation sector suggests a practical diagnostic for many-body Markov approximations: compare the shortest relevant system timescale with the longest environmental timescale. We find that the shortest relevant system timescale is $\tau_B$ in the single-excitation sector, which is both the fastest decay rate of the system and the emitted-pulse duration. Breaking the equivalency between decay rate and temporal width occurs with higher excitation numbers, but these are harder to analyze~\cite{Shi15,ArranzRegidor21,Lanuza24,Capurso26} or require particular assumptions about the environment~\cite{Vodenkova24,Windt25}. An analogous emergent non-Markovianity occurs due to collective strong coupling in cavity QED ~\cite{Guimond16,Burgess22,Das25,Gonzalez26}. Whether such a crossover occurs in free space is less clear. There, increasing emitter separations simultaneously lengthens propagation delays and weakens radiative interactions. Resolving how these competing effects govern memory will be important for engineering collective radiative interfaces~\cite{Rui20,Bohr24}, preserving quantum information in subradiant states~\cite{Guerin16,Ferioli21PRX,Douglas26arxiv}, and understanding the role of many-body superradiant error cascades in large-scale quantum computing devices~\cite{Mok24arxiv}.

\begin{acknowledgements}
\section*{Acknowledgements}
We thank Kanu Sinha and Ana Asenjo-Garcia for stimulating discussions. This research was supported in part by grant NSF PHY-2309135 to the Kavli Institute for Theoretical Physics (KITP).
\end{acknowledgements}

\bibliography{bibliography}

\begin{thebibliography}{73}%
\makeatletter
\providecommand \@ifxundefined [1]{%
 \@ifx{#1\undefined}
}%
\providecommand \@ifnum [1]{%
 \ifnum #1\expandafter \@firstoftwo
 \else \expandafter \@secondoftwo
 \fi
}%
\providecommand \@ifx [1]{%
 \ifx #1\expandafter \@firstoftwo
 \else \expandafter \@secondoftwo
 \fi
}%
\providecommand \natexlab [1]{#1}%
\providecommand \enquote  [1]{``#1''}%
\providecommand \bibnamefont  [1]{#1}%
\providecommand \bibfnamefont [1]{#1}%
\providecommand \citenamefont [1]{#1}%
\providecommand \href@noop [0]{\@secondoftwo}%
\providecommand \href [0]{\begingroup \@sanitize@url \@href}%
\providecommand \@href[1]{\@@startlink{#1}\@@href}%
\providecommand \@@href[1]{\endgroup#1\@@endlink}%
\providecommand \@sanitize@url [0]{\catcode `\\12\catcode `\$12\catcode `\&12\catcode `\#12\catcode `\^12\catcode `\_12\catcode `\%12\relax}%
\providecommand \@@startlink[1]{}%
\providecommand \@@endlink[0]{}%
\providecommand \url  [0]{\begingroup\@sanitize@url \@url }%
\providecommand \@url [1]{\endgroup\@href {#1}{\urlprefix }}%
\providecommand \urlprefix  [0]{URL }%
\providecommand \Eprint [0]{\href }%
\providecommand \doibase [0]{https://doi.org/}%
\providecommand \selectlanguage [0]{\@gobble}%
\providecommand \bibinfo  [0]{\@secondoftwo}%
\providecommand \bibfield  [0]{\@secondoftwo}%
\providecommand \translation [1]{[#1]}%
\providecommand \BibitemOpen [0]{}%
\providecommand \bibitemStop [0]{}%
\providecommand \bibitemNoStop [0]{.\EOS\space}%
\providecommand \EOS [0]{\spacefactor3000\relax}%
\providecommand \BibitemShut  [1]{\csname bibitem#1\endcsname}%
\let\auto@bib@innerbib\@empty
\bibitem [{\citenamefont {Breuer}\ and\ \citenamefont {Petruccione}(2007)}]{BreuerBook}%
  \BibitemOpen
  \bibfield  {author} {\bibinfo {author} {\bibfnamefont {H.-P.}\ \bibnamefont {Breuer}}\ and\ \bibinfo {author} {\bibfnamefont {F.}~\bibnamefont {Petruccione}},\ }\href {https://doi.org/10.1093/acprof:oso/9780199213900.001.0001} {\emph {\bibinfo {title} {The Theory of Open Quantum Systems}}}\ (\bibinfo  {publisher} {Oxford University Press},\ \bibinfo {year} {2007})\BibitemShut {NoStop}%
\bibitem [{\citenamefont {Rivas}\ and\ \citenamefont {Huelga}(2012)}]{RivasBook}%
  \BibitemOpen
  \bibfield  {author} {\bibinfo {author} {\bibfnamefont {A.}~\bibnamefont {Rivas}}\ and\ \bibinfo {author} {\bibfnamefont {S.~F.}\ \bibnamefont {Huelga}},\ }\href@noop {} {\emph {\bibinfo {title} {Open Quantum Systems: An Introduction}}},\ SpringerBriefs in Physics\ (\bibinfo  {publisher} {Springer Berlin, Heidelberg},\ \bibinfo {year} {2012})\BibitemShut {NoStop}%
\bibitem [{\citenamefont {Poyatos}\ \emph {et~al.}(1996)\citenamefont {Poyatos}, \citenamefont {Cirac},\ and\ \citenamefont {Zoller}}]{Poyatos96}%
  \BibitemOpen
  \bibfield  {author} {\bibinfo {author} {\bibfnamefont {J.~F.}\ \bibnamefont {Poyatos}}, \bibinfo {author} {\bibfnamefont {J.~I.}\ \bibnamefont {Cirac}},\ and\ \bibinfo {author} {\bibfnamefont {P.}~\bibnamefont {Zoller}},\ }\bibfield  {title} {\bibinfo {title} {Quantum reservoir engineering with laser cooled trapped ions},\ }\href {https://doi.org/10.1103/PhysRevLett.77.4728} {\bibfield  {journal} {\bibinfo  {journal} {Phys. Rev. Lett.}\ }\textbf {\bibinfo {volume} {77}},\ \bibinfo {pages} {4728} (\bibinfo {year} {1996})}\BibitemShut {NoStop}%
\bibitem [{\citenamefont {Cirac}\ \emph {et~al.}(1997)\citenamefont {Cirac}, \citenamefont {Zoller}, \citenamefont {Kimble},\ and\ \citenamefont {Mabuchi}}]{Cirac97}%
  \BibitemOpen
  \bibfield  {author} {\bibinfo {author} {\bibfnamefont {J.~I.}\ \bibnamefont {Cirac}}, \bibinfo {author} {\bibfnamefont {P.}~\bibnamefont {Zoller}}, \bibinfo {author} {\bibfnamefont {H.~J.}\ \bibnamefont {Kimble}},\ and\ \bibinfo {author} {\bibfnamefont {H.}~\bibnamefont {Mabuchi}},\ }\bibfield  {title} {\bibinfo {title} {Quantum state transfer and entanglement distribution among distant nodes in a quantum network},\ }\href {https://doi.org/10.1103/PhysRevLett.78.3221} {\bibfield  {journal} {\bibinfo  {journal} {Phys. Rev. Lett.}\ }\textbf {\bibinfo {volume} {78}},\ \bibinfo {pages} {3221} (\bibinfo {year} {1997})}\BibitemShut {NoStop}%
\bibitem [{\citenamefont {Verstraete}\ \emph {et~al.}(2009)\citenamefont {Verstraete}, \citenamefont {Wolf},\ and\ \citenamefont {Cirac}}]{Verstraete09}%
  \BibitemOpen
  \bibfield  {author} {\bibinfo {author} {\bibfnamefont {F.}~\bibnamefont {Verstraete}}, \bibinfo {author} {\bibfnamefont {M.~M.}\ \bibnamefont {Wolf}},\ and\ \bibinfo {author} {\bibfnamefont {J.~I.}\ \bibnamefont {Cirac}},\ }\bibfield  {title} {\bibinfo {title} {Quantum computation and quantum-state engineering driven by dissipation},\ }\href {https://doi.org/10.1038/nphys1342} {\bibfield  {journal} {\bibinfo  {journal} {Nat. Phys.}\ }\textbf {\bibinfo {volume} {5}},\ \bibinfo {pages} {633} (\bibinfo {year} {2009})}\BibitemShut {NoStop}%
\bibitem [{\citenamefont {Carmichael}(1993)}]{Carmichael93}%
  \BibitemOpen
  \bibfield  {author} {\bibinfo {author} {\bibfnamefont {H.~J.}\ \bibnamefont {Carmichael}},\ }\href@noop {} {\emph {\bibinfo {title} {An Open Systems Approach to Quantum Optics}}},\ \bibinfo {series} {Lecture Notes in Physics}, Vol.~\bibinfo {volume} {18}\ (\bibinfo  {publisher} {Springer-Verlag Berlin Heidelberg},\ \bibinfo {year} {1993})\BibitemShut {NoStop}%
\bibitem [{\citenamefont {Stefanini}\ \emph {et~al.}(2026)\citenamefont {Stefanini}, \citenamefont {Ziolkowska}, \citenamefont {Budker}, \citenamefont {Poschinger}, \citenamefont {Schmidt-Kaler}, \citenamefont {Browaeys}, \citenamefont {Imamoglu}, \citenamefont {Chang},\ and\ \citenamefont {Marino}}]{Stefanini26}%
  \BibitemOpen
  \bibfield  {author} {\bibinfo {author} {\bibfnamefont {M.}~\bibnamefont {Stefanini}}, \bibinfo {author} {\bibfnamefont {A.~A.}\ \bibnamefont {Ziolkowska}}, \bibinfo {author} {\bibfnamefont {D.}~\bibnamefont {Budker}}, \bibinfo {author} {\bibfnamefont {U.}~\bibnamefont {Poschinger}}, \bibinfo {author} {\bibfnamefont {F.}~\bibnamefont {Schmidt-Kaler}}, \bibinfo {author} {\bibfnamefont {A.}~\bibnamefont {Browaeys}}, \bibinfo {author} {\bibfnamefont {A.}~\bibnamefont {Imamoglu}}, \bibinfo {author} {\bibfnamefont {D.}~\bibnamefont {Chang}},\ and\ \bibinfo {author} {\bibfnamefont {J.}~\bibnamefont {Marino}},\ }\bibfield  {title} {\bibinfo {title} {Is {L}indblad for me?},\ }\href {https://doi.org/10.21468/SciPostPhysLectNotes.129} {\bibfield  {journal} {\bibinfo  {journal} {SciPost Phys. Lect. Notes}\ ,\ \bibinfo {pages} {129}} (\bibinfo {year} {2026})}\BibitemShut {NoStop}%
\bibitem [{\citenamefont {Dicke}(1954)}]{Dicke54}%
  \BibitemOpen
  \bibfield  {author} {\bibinfo {author} {\bibfnamefont {R.~H.}\ \bibnamefont {Dicke}},\ }\bibfield  {title} {\bibinfo {title} {Coherence in spontaneous radiation processes},\ }\href {https://doi.org/10.1103/PhysRev.93.99} {\bibfield  {journal} {\bibinfo  {journal} {Phys. Rev.}\ }\textbf {\bibinfo {volume} {93}},\ \bibinfo {pages} {99} (\bibinfo {year} {1954})}\BibitemShut {NoStop}%
\bibitem [{\citenamefont {Gross}\ and\ \citenamefont {Haroche}(1982)}]{Gross82}%
  \BibitemOpen
  \bibfield  {author} {\bibinfo {author} {\bibfnamefont {M.}~\bibnamefont {Gross}}\ and\ \bibinfo {author} {\bibfnamefont {S.}~\bibnamefont {Haroche}},\ }\bibfield  {title} {\bibinfo {title} {Superradiance: An essay on the theory of collective spontaneous emission},\ }\href {https://doi.org/https://doi.org/10.1016/0370-1573(82)90102-8} {\bibfield  {journal} {\bibinfo  {journal} {Phys. Rep.}\ }\textbf {\bibinfo {volume} {93}},\ \bibinfo {pages} {301} (\bibinfo {year} {1982})}\BibitemShut {NoStop}%
\bibitem [{\citenamefont {Benedict}\ \emph {et~al.}(1996)\citenamefont {Benedict}, \citenamefont {Ermolaev}, \citenamefont {Malyshev}, \citenamefont {Sokolov},\ and\ \citenamefont {Trifonov}}]{BenedictBook}%
  \BibitemOpen
  \bibfield  {author} {\bibinfo {author} {\bibfnamefont {M.~G.}\ \bibnamefont {Benedict}}, \bibinfo {author} {\bibfnamefont {A.~M.}\ \bibnamefont {Ermolaev}}, \bibinfo {author} {\bibfnamefont {V.~A.}\ \bibnamefont {Malyshev}}, \bibinfo {author} {\bibfnamefont {I.~V.}\ \bibnamefont {Sokolov}},\ and\ \bibinfo {author} {\bibfnamefont {E.~D.}\ \bibnamefont {Trifonov}},\ }\href@noop {} {\emph {\bibinfo {title} {Super-radiance: Multiatomic Coherent Emission}}}\ (\bibinfo  {publisher} {CRC Press},\ \bibinfo {year} {1996})\BibitemShut {NoStop}%
\bibitem [{\citenamefont {Masson}\ and\ \citenamefont {Asenjo-Garcia}(2022)}]{Masson22}%
  \BibitemOpen
  \bibfield  {author} {\bibinfo {author} {\bibfnamefont {S.~J.}\ \bibnamefont {Masson}}\ and\ \bibinfo {author} {\bibfnamefont {A.}~\bibnamefont {Asenjo-Garcia}},\ }\bibfield  {title} {\bibinfo {title} {Universality of {D}icke superradiance in arrays of quantum emitters},\ }\href {https://doi.org/10.1038/s41467-022-29805-4} {\bibfield  {journal} {\bibinfo  {journal} {Nat. Commun.}\ }\textbf {\bibinfo {volume} {13}},\ \bibinfo {pages} {2285} (\bibinfo {year} {2022})}\BibitemShut {NoStop}%
\bibitem [{\citenamefont {Palma}\ \emph {et~al.}(1996)\citenamefont {Palma}, \citenamefont {Suominen},\ and\ \citenamefont {Ekert}}]{Palma96}%
  \BibitemOpen
  \bibfield  {author} {\bibinfo {author} {\bibfnamefont {G.~M.}\ \bibnamefont {Palma}}, \bibinfo {author} {\bibfnamefont {K.-A.}\ \bibnamefont {Suominen}},\ and\ \bibinfo {author} {\bibfnamefont {A.}~\bibnamefont {Ekert}},\ }\bibfield  {title} {\bibinfo {title} {Quantum computers and dissipation},\ }\href {https://doi.org/10.1098/rspa.1996.0029} {\bibfield  {journal} {\bibinfo  {journal} {Proc. R. Soc. A: Math. Phys. Eng. Sci.}\ }\textbf {\bibinfo {volume} {452}},\ \bibinfo {pages} {567} (\bibinfo {year} {1996})}\BibitemShut {NoStop}%
\bibitem [{\citenamefont {Ramos}\ \emph {et~al.}(2014)\citenamefont {Ramos}, \citenamefont {Pichler}, \citenamefont {Daley},\ and\ \citenamefont {Zoller}}]{Ramos14}%
  \BibitemOpen
  \bibfield  {author} {\bibinfo {author} {\bibfnamefont {T.}~\bibnamefont {Ramos}}, \bibinfo {author} {\bibfnamefont {H.}~\bibnamefont {Pichler}}, \bibinfo {author} {\bibfnamefont {A.~J.}\ \bibnamefont {Daley}},\ and\ \bibinfo {author} {\bibfnamefont {P.}~\bibnamefont {Zoller}},\ }\bibfield  {title} {\bibinfo {title} {Quantum spin dimers from chiral dissipation in cold-atom chains},\ }\href {https://doi.org/10.1103/PhysRevLett.113.237203} {\bibfield  {journal} {\bibinfo  {journal} {Phys. Rev. Lett.}\ }\textbf {\bibinfo {volume} {113}},\ \bibinfo {pages} {237203} (\bibinfo {year} {2014})}\BibitemShut {NoStop}%
\bibitem [{\citenamefont {Asenjo-Garcia}\ \emph {et~al.}(2017)\citenamefont {Asenjo-Garcia}, \citenamefont {Moreno-Cardoner}, \citenamefont {Albrecht}, \citenamefont {Kimble},\ and\ \citenamefont {Chang}}]{Asenjo17PRX}%
  \BibitemOpen
  \bibfield  {author} {\bibinfo {author} {\bibfnamefont {A.}~\bibnamefont {Asenjo-Garcia}}, \bibinfo {author} {\bibfnamefont {M.}~\bibnamefont {Moreno-Cardoner}}, \bibinfo {author} {\bibfnamefont {A.}~\bibnamefont {Albrecht}}, \bibinfo {author} {\bibfnamefont {H.~J.}\ \bibnamefont {Kimble}},\ and\ \bibinfo {author} {\bibfnamefont {D.~E.}\ \bibnamefont {Chang}},\ }\bibfield  {title} {\bibinfo {title} {Exponential improvement in photon storage fidelities using subradiance and ``selective radiance'' in atomic arrays},\ }\href {https://doi.org/10.1103/PhysRevX.7.031024} {\bibfield  {journal} {\bibinfo  {journal} {Phys. Rev. X}\ }\textbf {\bibinfo {volume} {7}},\ \bibinfo {pages} {031024} (\bibinfo {year} {2017})}\BibitemShut {NoStop}%
\bibitem [{\citenamefont {Baumann}\ \emph {et~al.}(2010)\citenamefont {Baumann}, \citenamefont {Guerlin}, \citenamefont {Brennecke},\ and\ \citenamefont {Esslinger}}]{Baumann10}%
  \BibitemOpen
  \bibfield  {author} {\bibinfo {author} {\bibfnamefont {K.}~\bibnamefont {Baumann}}, \bibinfo {author} {\bibfnamefont {C.}~\bibnamefont {Guerlin}}, \bibinfo {author} {\bibfnamefont {F.}~\bibnamefont {Brennecke}},\ and\ \bibinfo {author} {\bibfnamefont {T.}~\bibnamefont {Esslinger}},\ }\bibfield  {title} {\bibinfo {title} {Dicke quantum phase transition with a superfluid gas in an optical cavity},\ }\href {https://doi.org/10.1038/nature09009} {\bibfield  {journal} {\bibinfo  {journal} {Nature}\ }\textbf {\bibinfo {volume} {464}},\ \bibinfo {pages} {1301} (\bibinfo {year} {2010})}\BibitemShut {NoStop}%
\bibitem [{\citenamefont {Fitzpatrick}\ \emph {et~al.}(2017)\citenamefont {Fitzpatrick}, \citenamefont {Sundaresan}, \citenamefont {Li}, \citenamefont {Koch},\ and\ \citenamefont {Houck}}]{Fitzpatrick17}%
  \BibitemOpen
  \bibfield  {author} {\bibinfo {author} {\bibfnamefont {M.}~\bibnamefont {Fitzpatrick}}, \bibinfo {author} {\bibfnamefont {N.~M.}\ \bibnamefont {Sundaresan}}, \bibinfo {author} {\bibfnamefont {A.~C.~Y.}\ \bibnamefont {Li}}, \bibinfo {author} {\bibfnamefont {J.}~\bibnamefont {Koch}},\ and\ \bibinfo {author} {\bibfnamefont {A.~A.}\ \bibnamefont {Houck}},\ }\bibfield  {title} {\bibinfo {title} {Observation of a dissipative phase transition in a one-dimensional circuit {QED} lattice},\ }\href {https://doi.org/10.1103/PhysRevX.7.011016} {\bibfield  {journal} {\bibinfo  {journal} {Phys. Rev. X}\ }\textbf {\bibinfo {volume} {7}},\ \bibinfo {pages} {011016} (\bibinfo {year} {2017})}\BibitemShut {NoStop}%
\bibitem [{\citenamefont {Benary}\ \emph {et~al.}(2022)\citenamefont {Benary}, \citenamefont {Baals}, \citenamefont {Bernhart}, \citenamefont {Jiang}, \citenamefont {R\"ohrle},\ and\ \citenamefont {Ott}}]{Benary22}%
  \BibitemOpen
  \bibfield  {author} {\bibinfo {author} {\bibfnamefont {J.}~\bibnamefont {Benary}}, \bibinfo {author} {\bibfnamefont {C.}~\bibnamefont {Baals}}, \bibinfo {author} {\bibfnamefont {E.}~\bibnamefont {Bernhart}}, \bibinfo {author} {\bibfnamefont {J.}~\bibnamefont {Jiang}}, \bibinfo {author} {\bibfnamefont {M.}~\bibnamefont {R\"ohrle}},\ and\ \bibinfo {author} {\bibfnamefont {H.}~\bibnamefont {Ott}},\ }\bibfield  {title} {\bibinfo {title} {Experimental observation of a dissipative phase transition in a multi-mode many-body quantum system},\ }\href {https://doi.org/10.1088/1367-2630/ac97b6} {\bibfield  {journal} {\bibinfo  {journal} {New J. Phys.}\ }\textbf {\bibinfo {volume} {24}},\ \bibinfo {pages} {103034} (\bibinfo {year} {2022})}\BibitemShut {NoStop}%
\bibitem [{\citenamefont {Arecchi}\ and\ \citenamefont {Courtens}(1970)}]{Arecchi70}%
  \BibitemOpen
  \bibfield  {author} {\bibinfo {author} {\bibfnamefont {F.~T.}\ \bibnamefont {Arecchi}}\ and\ \bibinfo {author} {\bibfnamefont {E.}~\bibnamefont {Courtens}},\ }\bibfield  {title} {\bibinfo {title} {Cooperative phenomena in resonant electromagnetic propagation},\ }\href {https://doi.org/10.1103/PhysRevA.2.1730} {\bibfield  {journal} {\bibinfo  {journal} {Phys. Rev. A}\ }\textbf {\bibinfo {volume} {2}},\ \bibinfo {pages} {1730} (\bibinfo {year} {1970})}\BibitemShut {NoStop}%
\bibitem [{\citenamefont {Bonifacio}\ and\ \citenamefont {Lugiato}(1975)}]{Bonifacio75}%
  \BibitemOpen
  \bibfield  {author} {\bibinfo {author} {\bibfnamefont {R.}~\bibnamefont {Bonifacio}}\ and\ \bibinfo {author} {\bibfnamefont {L.~A.}\ \bibnamefont {Lugiato}},\ }\bibfield  {title} {\bibinfo {title} {Cooperative radiation processes in two-level systems: Superfluorescence},\ }\href {https://doi.org/10.1103/PhysRevA.11.1507} {\bibfield  {journal} {\bibinfo  {journal} {Phys. Rev. A}\ }\textbf {\bibinfo {volume} {11}},\ \bibinfo {pages} {1507} (\bibinfo {year} {1975})}\BibitemShut {NoStop}%
\bibitem [{\citenamefont {Windt}\ \emph {et~al.}(2025)\citenamefont {Windt}, \citenamefont {Bello}, \citenamefont {Malz},\ and\ \citenamefont {Cirac}}]{Windt25}%
  \BibitemOpen
  \bibfield  {author} {\bibinfo {author} {\bibfnamefont {B.}~\bibnamefont {Windt}}, \bibinfo {author} {\bibfnamefont {M.}~\bibnamefont {Bello}}, \bibinfo {author} {\bibfnamefont {D.}~\bibnamefont {Malz}},\ and\ \bibinfo {author} {\bibfnamefont {J.~I.}\ \bibnamefont {Cirac}},\ }\bibfield  {title} {\bibinfo {title} {Effects of retardation on many-body superradiance in chiral waveguide {QED}},\ }\href {https://doi.org/10.1103/PhysRevLett.134.173601} {\bibfield  {journal} {\bibinfo  {journal} {Phys. Rev. Lett.}\ }\textbf {\bibinfo {volume} {134}},\ \bibinfo {pages} {173601} (\bibinfo {year} {2025})}\BibitemShut {NoStop}%
\bibitem [{\citenamefont {Zheng}\ and\ \citenamefont {Baranger}(2013)}]{Zheng13}%
  \BibitemOpen
  \bibfield  {author} {\bibinfo {author} {\bibfnamefont {H.}~\bibnamefont {Zheng}}\ and\ \bibinfo {author} {\bibfnamefont {H.~U.}\ \bibnamefont {Baranger}},\ }\bibfield  {title} {\bibinfo {title} {Persistent quantum beats and long-distance entanglement from waveguide-mediated interactions},\ }\href {https://doi.org/10.1103/PhysRevLett.110.113601} {\bibfield  {journal} {\bibinfo  {journal} {Phys. Rev. Lett.}\ }\textbf {\bibinfo {volume} {110}},\ \bibinfo {pages} {113601} (\bibinfo {year} {2013})}\BibitemShut {NoStop}%
\bibitem [{\citenamefont {Gonzalez-Ballestero}\ \emph {et~al.}(2013)\citenamefont {Gonzalez-Ballestero}, \citenamefont {Garc\'{i}a-Vidal},\ and\ \citenamefont {Moreno}}]{GonzalezBallestero13}%
  \BibitemOpen
  \bibfield  {author} {\bibinfo {author} {\bibfnamefont {C.}~\bibnamefont {Gonzalez-Ballestero}}, \bibinfo {author} {\bibfnamefont {F.~J.}\ \bibnamefont {Garc\'{i}a-Vidal}},\ and\ \bibinfo {author} {\bibfnamefont {E.}~\bibnamefont {Moreno}},\ }\bibfield  {title} {\bibinfo {title} {Non-{M}arkovian effects in waveguide-mediated entanglement},\ }\href {https://doi.org/10.1088/1367-2630/15/7/073015} {\bibfield  {journal} {\bibinfo  {journal} {New J. Phys.}\ }\textbf {\bibinfo {volume} {15}},\ \bibinfo {pages} {073015} (\bibinfo {year} {2013})}\BibitemShut {NoStop}%
\bibitem [{\citenamefont {Dinc}\ and\ \citenamefont {Bra\ifmmode~\acute{n}\else \'{n}\fi{}czyk}(2019)}]{Dinc19PRR}%
  \BibitemOpen
  \bibfield  {author} {\bibinfo {author} {\bibfnamefont {F.}~\bibnamefont {Dinc}}\ and\ \bibinfo {author} {\bibfnamefont {A.~M.}\ \bibnamefont {Bra\ifmmode~\acute{n}\else \'{n}\fi{}czyk}},\ }\bibfield  {title} {\bibinfo {title} {Non-{M}arkovian super-superradiance in a linear chain of up to 100 qubits},\ }\href {https://doi.org/10.1103/PhysRevResearch.1.032042} {\bibfield  {journal} {\bibinfo  {journal} {Phys. Rev. Res.}\ }\textbf {\bibinfo {volume} {1}},\ \bibinfo {pages} {032042(R)} (\bibinfo {year} {2019})}\BibitemShut {NoStop}%
\bibitem [{\citenamefont {Dinc}\ \emph {et~al.}(2019)\citenamefont {Dinc}, \citenamefont {Ercan},\ and\ \citenamefont {Bra{\'{n}}czyk}}]{Dinc19Quantum}%
  \BibitemOpen
  \bibfield  {author} {\bibinfo {author} {\bibfnamefont {F.}~\bibnamefont {Dinc}}, \bibinfo {author} {\bibfnamefont {{\.{I}}.}~\bibnamefont {Ercan}},\ and\ \bibinfo {author} {\bibfnamefont {A.~M.}\ \bibnamefont {Bra{\'{n}}czyk}},\ }\bibfield  {title} {\bibinfo {title} {Exact {M}arkovian and non-{M}arkovian time dynamics in waveguide {QED}: collective interactions, bound states in continuum, superradiance and subradiance},\ }\href {https://doi.org/10.22331/q-2019-12-09-213} {\bibfield  {journal} {\bibinfo  {journal} {{Quantum}}\ }\textbf {\bibinfo {volume} {3}},\ \bibinfo {pages} {213} (\bibinfo {year} {2019})}\BibitemShut {NoStop}%
\bibitem [{\citenamefont {Sinha}\ \emph {et~al.}(2020{\natexlab{a}})\citenamefont {Sinha}, \citenamefont {Meystre}, \citenamefont {Goldschmidt}, \citenamefont {Fatemi}, \citenamefont {Rolston},\ and\ \citenamefont {Solano}}]{Sinha20}%
  \BibitemOpen
  \bibfield  {author} {\bibinfo {author} {\bibfnamefont {K.}~\bibnamefont {Sinha}}, \bibinfo {author} {\bibfnamefont {P.}~\bibnamefont {Meystre}}, \bibinfo {author} {\bibfnamefont {E.~A.}\ \bibnamefont {Goldschmidt}}, \bibinfo {author} {\bibfnamefont {F.~K.}\ \bibnamefont {Fatemi}}, \bibinfo {author} {\bibfnamefont {S.~L.}\ \bibnamefont {Rolston}},\ and\ \bibinfo {author} {\bibfnamefont {P.}~\bibnamefont {Solano}},\ }\bibfield  {title} {\bibinfo {title} {Non-{M}arkovian collective emission from macroscopically separated emitters},\ }\href {https://doi.org/10.1103/PhysRevLett.124.043603} {\bibfield  {journal} {\bibinfo  {journal} {Phys. Rev. Lett.}\ }\textbf {\bibinfo {volume} {124}},\ \bibinfo {pages} {043603} (\bibinfo {year} {2020}{\natexlab{a}})}\BibitemShut {NoStop}%
\bibitem [{\citenamefont {Barahona-Pascual}\ \emph {et~al.}(2025)\citenamefont {Barahona-Pascual}, \citenamefont {Jiang}, \citenamefont {Santos},\ and\ \citenamefont {Garc\'ia-Ripoll}}]{BarahonaPascual25}%
  \BibitemOpen
  \bibfield  {author} {\bibinfo {author} {\bibfnamefont {C.}~\bibnamefont {Barahona-Pascual}}, \bibinfo {author} {\bibfnamefont {H.}~\bibnamefont {Jiang}}, \bibinfo {author} {\bibfnamefont {A.~C.}\ \bibnamefont {Santos}},\ and\ \bibinfo {author} {\bibfnamefont {J.~J.}\ \bibnamefont {Garc\'ia-Ripoll}},\ }\bibfield  {title} {\bibinfo {title} {Time-delayed collective dynamics in waveguide {QED} and bosonic quantum networks},\ }\href {https://doi.org/10.1088/2058-9565/ae2291} {\bibfield  {journal} {\bibinfo  {journal} {Quantum Sci. Technol.}\ }\textbf {\bibinfo {volume} {11}},\ \bibinfo {pages} {015016} (\bibinfo {year} {2025})}\BibitemShut {NoStop}%
\bibitem [{\citenamefont {Capurso}\ \emph {et~al.}(2026)\citenamefont {Capurso}, \citenamefont {Calaj\'o}, \citenamefont {Montangero}, \citenamefont {Pascazio}, \citenamefont {Pepe}, \citenamefont {Maffei}, \citenamefont {Magnifico},\ and\ \citenamefont {Facchi}}]{Capurso26}%
  \BibitemOpen
  \bibfield  {author} {\bibinfo {author} {\bibfnamefont {R.~L.}\ \bibnamefont {Capurso}}, \bibinfo {author} {\bibfnamefont {G.}~\bibnamefont {Calaj\'o}}, \bibinfo {author} {\bibfnamefont {S.}~\bibnamefont {Montangero}}, \bibinfo {author} {\bibfnamefont {S.}~\bibnamefont {Pascazio}}, \bibinfo {author} {\bibfnamefont {F.~V.}\ \bibnamefont {Pepe}}, \bibinfo {author} {\bibfnamefont {M.}~\bibnamefont {Maffei}}, \bibinfo {author} {\bibfnamefont {G.}~\bibnamefont {Magnifico}},\ and\ \bibinfo {author} {\bibfnamefont {P.}~\bibnamefont {Facchi}},\ }\bibfield  {title} {\bibinfo {title} {Superradiant decay in non-{M}arkovian waveguide quantum electrodynamics},\ }\bibfield  {journal} {\bibinfo  {journal} {Commun. Phys.}\ }\href {https://doi.org/10.1038/s42005-026-02722-4} {10.1038/s42005-026-02722-4} (\bibinfo {year} {2026})\BibitemShut {NoStop}%
\bibitem [{\citenamefont {Calaj\'o}\ \emph {et~al.}(2019)\citenamefont {Calaj\'o}, \citenamefont {Fang}, \citenamefont {Baranger},\ and\ \citenamefont {Ciccarello}}]{Calajo19}%
  \BibitemOpen
  \bibfield  {author} {\bibinfo {author} {\bibfnamefont {G.}~\bibnamefont {Calaj\'o}}, \bibinfo {author} {\bibfnamefont {Y.-L.~L.}\ \bibnamefont {Fang}}, \bibinfo {author} {\bibfnamefont {H.~U.}\ \bibnamefont {Baranger}},\ and\ \bibinfo {author} {\bibfnamefont {F.}~\bibnamefont {Ciccarello}},\ }\bibfield  {title} {\bibinfo {title} {Exciting a bound state in the continuum through multiphoton scattering plus delayed quantum feedback},\ }\href {https://doi.org/10.1103/PhysRevLett.122.073601} {\bibfield  {journal} {\bibinfo  {journal} {Phys. Rev. Lett.}\ }\textbf {\bibinfo {volume} {122}},\ \bibinfo {pages} {073601} (\bibinfo {year} {2019})}\BibitemShut {NoStop}%
\bibitem [{\citenamefont {Carmele}\ \emph {et~al.}(2020)\citenamefont {Carmele}, \citenamefont {Nemet}, \citenamefont {Canela},\ and\ \citenamefont {Parkins}}]{Carmele20}%
  \BibitemOpen
  \bibfield  {author} {\bibinfo {author} {\bibfnamefont {A.}~\bibnamefont {Carmele}}, \bibinfo {author} {\bibfnamefont {N.}~\bibnamefont {Nemet}}, \bibinfo {author} {\bibfnamefont {V.}~\bibnamefont {Canela}},\ and\ \bibinfo {author} {\bibfnamefont {S.}~\bibnamefont {Parkins}},\ }\bibfield  {title} {\bibinfo {title} {Pronounced non-{M}arkovian features in multiply excited, multiple emitter waveguide {QED}: Retardation induced anomalous population trapping},\ }\href {https://doi.org/10.1103/PhysRevResearch.2.013238} {\bibfield  {journal} {\bibinfo  {journal} {Phys. Rev. Research}\ }\textbf {\bibinfo {volume} {2}},\ \bibinfo {pages} {013238} (\bibinfo {year} {2020})}\BibitemShut {NoStop}%
\bibitem [{\citenamefont {Alvarez-Giron}\ \emph {et~al.}(2024)\citenamefont {Alvarez-Giron}, \citenamefont {Solano}, \citenamefont {Sinha},\ and\ \citenamefont {Barberis-Blostein}}]{AlvarezGiron24}%
  \BibitemOpen
  \bibfield  {author} {\bibinfo {author} {\bibfnamefont {W.}~\bibnamefont {Alvarez-Giron}}, \bibinfo {author} {\bibfnamefont {P.}~\bibnamefont {Solano}}, \bibinfo {author} {\bibfnamefont {K.}~\bibnamefont {Sinha}},\ and\ \bibinfo {author} {\bibfnamefont {P.}~\bibnamefont {Barberis-Blostein}},\ }\bibfield  {title} {\bibinfo {title} {Delay-induced spontaneous dark-state generation from two distant excited atoms},\ }\href {https://doi.org/10.1103/PhysRevResearch.6.023213} {\bibfield  {journal} {\bibinfo  {journal} {Phys. Rev. Res.}\ }\textbf {\bibinfo {volume} {6}},\ \bibinfo {pages} {023213} (\bibinfo {year} {2024})}\BibitemShut {NoStop}%
\bibitem [{\citenamefont {Magnifico}\ \emph {et~al.}(2025)\citenamefont {Magnifico}, \citenamefont {Maffei}, \citenamefont {Pomarico}, \citenamefont {Das}, \citenamefont {Facchi}, \citenamefont {Pascazio},\ and\ \citenamefont {Pepe}}]{Magnifico25}%
  \BibitemOpen
  \bibfield  {author} {\bibinfo {author} {\bibfnamefont {G.}~\bibnamefont {Magnifico}}, \bibinfo {author} {\bibfnamefont {M.}~\bibnamefont {Maffei}}, \bibinfo {author} {\bibfnamefont {D.}~\bibnamefont {Pomarico}}, \bibinfo {author} {\bibfnamefont {D.}~\bibnamefont {Das}}, \bibinfo {author} {\bibfnamefont {P.}~\bibnamefont {Facchi}}, \bibinfo {author} {\bibfnamefont {S.}~\bibnamefont {Pascazio}},\ and\ \bibinfo {author} {\bibfnamefont {F.~V.}\ \bibnamefont {Pepe}},\ }\bibfield  {title} {\bibinfo {title} {Non-{M}arkovian dynamics of generation of bound states in the continuum via single-photon scattering},\ }\href {https://doi.org/10.1103/gtf6-zb57} {\bibfield  {journal} {\bibinfo  {journal} {Phys. Rev. Res.}\ }\textbf {\bibinfo {volume} {7}},\ \bibinfo {pages} {033249} (\bibinfo {year} {2025})}\BibitemShut {NoStop}%
\bibitem [{\citenamefont {Guha}\ \emph {et~al.}(2025)\citenamefont {Guha}, \citenamefont {Bar}, \citenamefont {Agarwalla},\ and\ \citenamefont {Venkatesh}}]{Guha25}%
  \BibitemOpen
  \bibfield  {author} {\bibinfo {author} {\bibfnamefont {S.}~\bibnamefont {Guha}}, \bibinfo {author} {\bibfnamefont {I.}~\bibnamefont {Bar}}, \bibinfo {author} {\bibfnamefont {B.~K.}\ \bibnamefont {Agarwalla}},\ and\ \bibinfo {author} {\bibfnamefont {B.~P.}\ \bibnamefont {Venkatesh}},\ }\bibfield  {title} {\bibinfo {title} {Collective dissipation of oscillator dipoles strongly coupled to one-dimensional electromagnetic reservoirs},\ }\href {https://doi.org/10.1103/dnnn-5b5p} {\bibfield  {journal} {\bibinfo  {journal} {Phys. Rev. A}\ }\textbf {\bibinfo {volume} {112}},\ \bibinfo {pages} {043709} (\bibinfo {year} {2025})}\BibitemShut {NoStop}%
\bibitem [{\citenamefont {Vera}\ \emph {et~al.}(2026)\citenamefont {Vera}, \citenamefont {Quinteros}, \citenamefont {Barberis-Blostein},\ and\ \citenamefont {Solano}}]{Vera26arxiv}%
  \BibitemOpen
  \bibfield  {author} {\bibinfo {author} {\bibfnamefont {N.}~\bibnamefont {Vera}}, \bibinfo {author} {\bibfnamefont {F.~M.}\ \bibnamefont {Quinteros}}, \bibinfo {author} {\bibfnamefont {P.}~\bibnamefont {Barberis-Blostein}},\ and\ \bibinfo {author} {\bibfnamefont {P.}~\bibnamefont {Solano}},\ }\bibfield  {title} {\bibinfo {title} {Correlation localization in waveguide {QED} with delayed interactions},\ }\href@noop {} {\bibfield  {journal} {\bibinfo  {journal} {arXiv:2607.06888}\ } (\bibinfo {year} {2026})}\BibitemShut {NoStop}%
\bibitem [{\citenamefont {Mirhosseini}\ \emph {et~al.}(2018)\citenamefont {Mirhosseini}, \citenamefont {Kim}, \citenamefont {Ferreira}, \citenamefont {Kalaee}, \citenamefont {Sipahigil}, \citenamefont {Keller},\ and\ \citenamefont {Painter}}]{Mirhosseini18}%
  \BibitemOpen
  \bibfield  {author} {\bibinfo {author} {\bibfnamefont {M.}~\bibnamefont {Mirhosseini}}, \bibinfo {author} {\bibfnamefont {E.}~\bibnamefont {Kim}}, \bibinfo {author} {\bibfnamefont {V.~S.}\ \bibnamefont {Ferreira}}, \bibinfo {author} {\bibfnamefont {M.}~\bibnamefont {Kalaee}}, \bibinfo {author} {\bibfnamefont {A.}~\bibnamefont {Sipahigil}}, \bibinfo {author} {\bibfnamefont {A.~J.}\ \bibnamefont {Keller}},\ and\ \bibinfo {author} {\bibfnamefont {O.}~\bibnamefont {Painter}},\ }\bibfield  {title} {\bibinfo {title} {Superconducting metamaterials for waveguide quantum electrodynamics},\ }\href {https://doi.org/10.1038/s41467-018-06142-z} {\bibfield  {journal} {\bibinfo  {journal} {Nat. Commun.}\ }\textbf {\bibinfo {volume} {9}},\ \bibinfo {pages} {3706} (\bibinfo {year} {2018})}\BibitemShut {NoStop}%
\bibitem [{\citenamefont {Scigliuzzo}\ \emph {et~al.}(2022)\citenamefont {Scigliuzzo}, \citenamefont {Calaj\`o}, \citenamefont {Ciccarello}, \citenamefont {Perez~Lozano}, \citenamefont {Bengtsson}, \citenamefont {Scarlino}, \citenamefont {Wallraff}, \citenamefont {Chang}, \citenamefont {Delsing},\ and\ \citenamefont {Gasparinetti}}]{Scigliuzzo22}%
  \BibitemOpen
  \bibfield  {author} {\bibinfo {author} {\bibfnamefont {M.}~\bibnamefont {Scigliuzzo}}, \bibinfo {author} {\bibfnamefont {G.}~\bibnamefont {Calaj\`o}}, \bibinfo {author} {\bibfnamefont {F.}~\bibnamefont {Ciccarello}}, \bibinfo {author} {\bibfnamefont {D.}~\bibnamefont {Perez~Lozano}}, \bibinfo {author} {\bibfnamefont {A.}~\bibnamefont {Bengtsson}}, \bibinfo {author} {\bibfnamefont {P.}~\bibnamefont {Scarlino}}, \bibinfo {author} {\bibfnamefont {A.}~\bibnamefont {Wallraff}}, \bibinfo {author} {\bibfnamefont {D.}~\bibnamefont {Chang}}, \bibinfo {author} {\bibfnamefont {P.}~\bibnamefont {Delsing}},\ and\ \bibinfo {author} {\bibfnamefont {S.}~\bibnamefont {Gasparinetti}},\ }\bibfield  {title} {\bibinfo {title} {Controlling atom-photon bound states in an array of {J}osephson-junction resonators},\ }\href {https://doi.org/10.1103/PhysRevX.12.031036} {\bibfield  {journal} {\bibinfo  {journal} {Phys. Rev. X}\ }\textbf {\bibinfo {volume} {12}},\ \bibinfo {pages} {031036} (\bibinfo {year} {2022})}\BibitemShut
  {NoStop}%
\bibitem [{\citenamefont {Goban}\ \emph {et~al.}(2015)\citenamefont {Goban}, \citenamefont {Hung}, \citenamefont {Hood}, \citenamefont {Yu}, \citenamefont {Muniz}, \citenamefont {Painter},\ and\ \citenamefont {Kimble}}]{Goban15}%
  \BibitemOpen
  \bibfield  {author} {\bibinfo {author} {\bibfnamefont {A.}~\bibnamefont {Goban}}, \bibinfo {author} {\bibfnamefont {C.-L.}\ \bibnamefont {Hung}}, \bibinfo {author} {\bibfnamefont {J.~D.}\ \bibnamefont {Hood}}, \bibinfo {author} {\bibfnamefont {S.-P.}\ \bibnamefont {Yu}}, \bibinfo {author} {\bibfnamefont {J.~A.}\ \bibnamefont {Muniz}}, \bibinfo {author} {\bibfnamefont {O.}~\bibnamefont {Painter}},\ and\ \bibinfo {author} {\bibfnamefont {H.~J.}\ \bibnamefont {Kimble}},\ }\bibfield  {title} {\bibinfo {title} {Superradiance for atoms trapped along a photonic crystal waveguide},\ }\href {https://doi.org/10.1103/PhysRevLett.115.063601} {\bibfield  {journal} {\bibinfo  {journal} {Phys. Rev. Lett.}\ }\textbf {\bibinfo {volume} {115}},\ \bibinfo {pages} {063601} (\bibinfo {year} {2015})}\BibitemShut {NoStop}%
\bibitem [{\citenamefont {Tiranov}\ \emph {et~al.}(2023)\citenamefont {Tiranov}, \citenamefont {Angelopoulou}, \citenamefont {van Diepen}, \citenamefont {Schrinski}, \citenamefont {Sandberg}, \citenamefont {Wang}, \citenamefont {Midolo}, \citenamefont {Scholz}, \citenamefont {Wieck}, \citenamefont {Ludwig}, \citenamefont {S{\o}rensen},\ and\ \citenamefont {Lodahl}}]{Tiranov23}%
  \BibitemOpen
  \bibfield  {author} {\bibinfo {author} {\bibfnamefont {A.}~\bibnamefont {Tiranov}}, \bibinfo {author} {\bibfnamefont {V.}~\bibnamefont {Angelopoulou}}, \bibinfo {author} {\bibfnamefont {C.~J.}\ \bibnamefont {van Diepen}}, \bibinfo {author} {\bibfnamefont {B.}~\bibnamefont {Schrinski}}, \bibinfo {author} {\bibfnamefont {O.~A.~D.}\ \bibnamefont {Sandberg}}, \bibinfo {author} {\bibfnamefont {Y.}~\bibnamefont {Wang}}, \bibinfo {author} {\bibfnamefont {L.}~\bibnamefont {Midolo}}, \bibinfo {author} {\bibfnamefont {S.}~\bibnamefont {Scholz}}, \bibinfo {author} {\bibfnamefont {A.~D.}\ \bibnamefont {Wieck}}, \bibinfo {author} {\bibfnamefont {A.}~\bibnamefont {Ludwig}}, \bibinfo {author} {\bibfnamefont {A.~S.}\ \bibnamefont {S{\o}rensen}},\ and\ \bibinfo {author} {\bibfnamefont {P.}~\bibnamefont {Lodahl}},\ }\bibfield  {title} {\bibinfo {title} {Collective super- and subradiant dynamics between distant optical quantum emitters},\ }\href {https://doi.org/10.1126/science.ade9324} {\bibfield  {journal} {\bibinfo
  {journal} {Science}\ }\textbf {\bibinfo {volume} {379}},\ \bibinfo {pages} {389} (\bibinfo {year} {2023})}\BibitemShut {NoStop}%
\bibitem [{\citenamefont {Gustafsson}\ \emph {et~al.}(2014)\citenamefont {Gustafsson}, \citenamefont {Aref}, \citenamefont {Kockum}, \citenamefont {Ekstr\"om}, \citenamefont {Johansson},\ and\ \citenamefont {Delsing}}]{Gustafsson14}%
  \BibitemOpen
  \bibfield  {author} {\bibinfo {author} {\bibfnamefont {M.~V.}\ \bibnamefont {Gustafsson}}, \bibinfo {author} {\bibfnamefont {T.}~\bibnamefont {Aref}}, \bibinfo {author} {\bibfnamefont {A.~F.}\ \bibnamefont {Kockum}}, \bibinfo {author} {\bibfnamefont {M.~K.}\ \bibnamefont {Ekstr\"om}}, \bibinfo {author} {\bibfnamefont {G.}~\bibnamefont {Johansson}},\ and\ \bibinfo {author} {\bibfnamefont {P.}~\bibnamefont {Delsing}},\ }\bibfield  {title} {\bibinfo {title} {Propagating phonons coupled to an artificial atom},\ }\href {https://doi.org/10.1126/science.1257219} {\bibfield  {journal} {\bibinfo  {journal} {Science}\ }\textbf {\bibinfo {volume} {346}},\ \bibinfo {pages} {207} (\bibinfo {year} {2014})}\BibitemShut {NoStop}%
\bibitem [{\citenamefont {Dumur}\ \emph {et~al.}(2021)\citenamefont {Dumur}, \citenamefont {Satzinger}, \citenamefont {Peairs}, \citenamefont {Chou}, \citenamefont {Bienfait}, \citenamefont {Chang}, \citenamefont {Conner}, \citenamefont {Grebel}, \citenamefont {Povey}, \citenamefont {Zhong},\ and\ \citenamefont {Cleland}}]{Dumur21}%
  \BibitemOpen
  \bibfield  {author} {\bibinfo {author} {\bibfnamefont {{\'E}.}~\bibnamefont {Dumur}}, \bibinfo {author} {\bibfnamefont {K.~J.}\ \bibnamefont {Satzinger}}, \bibinfo {author} {\bibfnamefont {G.~A.}\ \bibnamefont {Peairs}}, \bibinfo {author} {\bibfnamefont {M.~H.}\ \bibnamefont {Chou}}, \bibinfo {author} {\bibfnamefont {A.}~\bibnamefont {Bienfait}}, \bibinfo {author} {\bibfnamefont {H.~S.}\ \bibnamefont {Chang}}, \bibinfo {author} {\bibfnamefont {C.~R.}\ \bibnamefont {Conner}}, \bibinfo {author} {\bibfnamefont {J.}~\bibnamefont {Grebel}}, \bibinfo {author} {\bibfnamefont {R.~G.}\ \bibnamefont {Povey}}, \bibinfo {author} {\bibfnamefont {Y.~P.}\ \bibnamefont {Zhong}},\ and\ \bibinfo {author} {\bibfnamefont {A.~N.}\ \bibnamefont {Cleland}},\ }\bibfield  {title} {\bibinfo {title} {Quantum communication with itinerant surface acoustic wave phonons},\ }\href {https://doi.org/10.1038/s41534-021-00511-1} {\bibfield  {journal} {\bibinfo  {journal} {npj Quantum Inf.}\ }\textbf {\bibinfo {volume} {7}},\ \bibinfo {pages}
  {173} (\bibinfo {year} {2021})}\BibitemShut {NoStop}%
\bibitem [{\citenamefont {Krinner}\ \emph {et~al.}(2018)\citenamefont {Krinner}, \citenamefont {Stewart}, \citenamefont {Pazmi\~no}, \citenamefont {Kwon},\ and\ \citenamefont {Schneble}}]{Krinner18}%
  \BibitemOpen
  \bibfield  {author} {\bibinfo {author} {\bibfnamefont {L.}~\bibnamefont {Krinner}}, \bibinfo {author} {\bibfnamefont {M.}~\bibnamefont {Stewart}}, \bibinfo {author} {\bibfnamefont {A.}~\bibnamefont {Pazmi\~no}}, \bibinfo {author} {\bibfnamefont {J.}~\bibnamefont {Kwon}},\ and\ \bibinfo {author} {\bibfnamefont {D.}~\bibnamefont {Schneble}},\ }\bibfield  {title} {\bibinfo {title} {Spontaneous emission of matter waves from a tunable open quantum system},\ }\href {https://doi.org/10.1038/s41586-018-0348-z} {\bibfield  {journal} {\bibinfo  {journal} {Nature}\ }\textbf {\bibinfo {volume} {559}},\ \bibinfo {pages} {589} (\bibinfo {year} {2018})}\BibitemShut {NoStop}%
\bibitem [{\citenamefont {Kim}\ \emph {et~al.}(2025)\citenamefont {Kim}, \citenamefont {Lanuza},\ and\ \citenamefont {Schneble}}]{Kim25}%
  \BibitemOpen
  \bibfield  {author} {\bibinfo {author} {\bibfnamefont {Y.}~\bibnamefont {Kim}}, \bibinfo {author} {\bibfnamefont {A.}~\bibnamefont {Lanuza}},\ and\ \bibinfo {author} {\bibfnamefont {D.}~\bibnamefont {Schneble}},\ }\bibfield  {title} {\bibinfo {title} {Super- and subradiant dynamics of quantum emitters mediated by atomic matter waves},\ }\href {https://doi.org/10.1038/s41567-024-02676-w} {\bibfield  {journal} {\bibinfo  {journal} {Nat. Phys.}\ }\textbf {\bibinfo {volume} {21}},\ \bibinfo {pages} {70} (\bibinfo {year} {2025})}\BibitemShut {NoStop}%
\bibitem [{\citenamefont {Lodahl}\ \emph {et~al.}(2015)\citenamefont {Lodahl}, \citenamefont {Mahmoodian},\ and\ \citenamefont {Stobbe}}]{Lodahl15}%
  \BibitemOpen
  \bibfield  {author} {\bibinfo {author} {\bibfnamefont {P.}~\bibnamefont {Lodahl}}, \bibinfo {author} {\bibfnamefont {S.}~\bibnamefont {Mahmoodian}},\ and\ \bibinfo {author} {\bibfnamefont {S.}~\bibnamefont {Stobbe}},\ }\bibfield  {title} {\bibinfo {title} {Interfacing single photons and single quantum dots with photonic nanostructures},\ }\href {https://doi.org/10.1103/RevModPhys.87.347} {\bibfield  {journal} {\bibinfo  {journal} {Rev. Mod. Phys.}\ }\textbf {\bibinfo {volume} {87}},\ \bibinfo {pages} {347} (\bibinfo {year} {2015})}\BibitemShut {NoStop}%
\bibitem [{\citenamefont {Wang}\ \emph {et~al.}(2018)\citenamefont {Wang}, \citenamefont {Chernikov}, \citenamefont {Glazov}, \citenamefont {Heinz}, \citenamefont {Marie}, \citenamefont {Amand},\ and\ \citenamefont {Urbaszek}}]{Wang18}%
  \BibitemOpen
  \bibfield  {author} {\bibinfo {author} {\bibfnamefont {G.}~\bibnamefont {Wang}}, \bibinfo {author} {\bibfnamefont {A.}~\bibnamefont {Chernikov}}, \bibinfo {author} {\bibfnamefont {M.~M.}\ \bibnamefont {Glazov}}, \bibinfo {author} {\bibfnamefont {T.~F.}\ \bibnamefont {Heinz}}, \bibinfo {author} {\bibfnamefont {X.}~\bibnamefont {Marie}}, \bibinfo {author} {\bibfnamefont {T.}~\bibnamefont {Amand}},\ and\ \bibinfo {author} {\bibfnamefont {B.}~\bibnamefont {Urbaszek}},\ }\bibfield  {title} {\bibinfo {title} {Colloquium: Excitons in atomically thin transition metal dichalcogenides},\ }\href {https://doi.org/10.1103/RevModPhys.90.021001} {\bibfield  {journal} {\bibinfo  {journal} {Rev. Mod. Phys.}\ }\textbf {\bibinfo {volume} {90}},\ \bibinfo {pages} {021001} (\bibinfo {year} {2018})}\BibitemShut {NoStop}%
\bibitem [{\citenamefont {Feldmann}\ \emph {et~al.}(1987)\citenamefont {Feldmann}, \citenamefont {Peter}, \citenamefont {G\"obel}, \citenamefont {Dawson}, \citenamefont {Moore}, \citenamefont {Foxon},\ and\ \citenamefont {Elliott}}]{Feldmann87}%
  \BibitemOpen
  \bibfield  {author} {\bibinfo {author} {\bibfnamefont {J.}~\bibnamefont {Feldmann}}, \bibinfo {author} {\bibfnamefont {G.}~\bibnamefont {Peter}}, \bibinfo {author} {\bibfnamefont {E.~O.}\ \bibnamefont {G\"obel}}, \bibinfo {author} {\bibfnamefont {P.}~\bibnamefont {Dawson}}, \bibinfo {author} {\bibfnamefont {K.}~\bibnamefont {Moore}}, \bibinfo {author} {\bibfnamefont {C.}~\bibnamefont {Foxon}},\ and\ \bibinfo {author} {\bibfnamefont {R.~J.}\ \bibnamefont {Elliott}},\ }\bibfield  {title} {\bibinfo {title} {Linewidth dependence of radiative exciton lifetimes in quantum wells},\ }\href {https://doi.org/10.1103/PhysRevLett.59.2337} {\bibfield  {journal} {\bibinfo  {journal} {Phys. Rev. Lett.}\ }\textbf {\bibinfo {volume} {59}},\ \bibinfo {pages} {2337} (\bibinfo {year} {1987})}\BibitemShut {NoStop}%
\bibitem [{Si_()}]{Si_enm}%
  \BibitemOpen
  \href@noop {} {}\bibinfo {note} {See Supplemental Material for analytic justification of the role of $\chi$, discussions of the role of imperfect waveguide coupling, chiral coupling, and arbitrary separation, and discussion of how the intensity and power spectra are calculated, including Refs.~\cite{CardenasLopez23,ScullyBook}.}\BibitemShut {Stop}%
\bibitem [{\citenamefont {Chang}\ \emph {et~al.}(2012)\citenamefont {Chang}, \citenamefont {Jiang}, \citenamefont {Gorshkov},\ and\ \citenamefont {Kimble}}]{Chang12}%
  \BibitemOpen
  \bibfield  {author} {\bibinfo {author} {\bibfnamefont {D.~E.}\ \bibnamefont {Chang}}, \bibinfo {author} {\bibfnamefont {L.}~\bibnamefont {Jiang}}, \bibinfo {author} {\bibfnamefont {A.~V.}\ \bibnamefont {Gorshkov}},\ and\ \bibinfo {author} {\bibfnamefont {H.~J.}\ \bibnamefont {Kimble}},\ }\bibfield  {title} {\bibinfo {title} {Cavity {QED} with atomic mirrors},\ }\href {http://stacks.iop.org/1367-2630/14/i=6/a=063003} {\bibfield  {journal} {\bibinfo  {journal} {New J. Phys.}\ }\textbf {\bibinfo {volume} {14}},\ \bibinfo {pages} {063003} (\bibinfo {year} {2012})}\BibitemShut {NoStop}%
\bibitem [{\citenamefont {Sinha}\ \emph {et~al.}(2020{\natexlab{b}})\citenamefont {Sinha}, \citenamefont {Gonz\'alez-Tudela}, \citenamefont {Lu},\ and\ \citenamefont {Solano}}]{Sinha20PRA}%
  \BibitemOpen
  \bibfield  {author} {\bibinfo {author} {\bibfnamefont {K.}~\bibnamefont {Sinha}}, \bibinfo {author} {\bibfnamefont {A.}~\bibnamefont {Gonz\'alez-Tudela}}, \bibinfo {author} {\bibfnamefont {Y.}~\bibnamefont {Lu}},\ and\ \bibinfo {author} {\bibfnamefont {P.}~\bibnamefont {Solano}},\ }\bibfield  {title} {\bibinfo {title} {Collective radiation from distant emitters},\ }\href {https://doi.org/10.1103/PhysRevA.102.043718} {\bibfield  {journal} {\bibinfo  {journal} {Phys. Rev. A}\ }\textbf {\bibinfo {volume} {102}},\ \bibinfo {pages} {043718} (\bibinfo {year} {2020}{\natexlab{b}})}\BibitemShut {NoStop}%
\bibitem [{\citenamefont {Cilluffo}\ \emph {et~al.}(2026)\citenamefont {Cilluffo}, \citenamefont {Ferialdi}, \citenamefont {Palma}, \citenamefont {Calaj\`o},\ and\ \citenamefont {Ciccarello}}]{Cilluffo26}%
  \BibitemOpen
  \bibfield  {author} {\bibinfo {author} {\bibfnamefont {D.}~\bibnamefont {Cilluffo}}, \bibinfo {author} {\bibfnamefont {L.}~\bibnamefont {Ferialdi}}, \bibinfo {author} {\bibfnamefont {G.~M.}\ \bibnamefont {Palma}}, \bibinfo {author} {\bibfnamefont {G.}~\bibnamefont {Calaj\`o}},\ and\ \bibinfo {author} {\bibfnamefont {F.}~\bibnamefont {Ciccarello}},\ }\bibfield  {title} {\bibinfo {title} {Multimode-cavity picture of non-{M}arkovian waveguide {QED}},\ }\href {https://doi.org/10.1103/hc4r-2st8} {\bibfield  {journal} {\bibinfo  {journal} {Phys. Rev. Res.}\ }\textbf {\bibinfo {volume} {8}},\ \bibinfo {pages} {023172} (\bibinfo {year} {2026})}\BibitemShut {NoStop}%
\bibitem [{\citenamefont {Dhara}\ \emph {et~al.}(2026)\citenamefont {Dhara}, \citenamefont {Padilla}, \citenamefont {Guha}, \citenamefont {Das},\ and\ \citenamefont {Sinha}}]{Dhara26arxiv}%
  \BibitemOpen
  \bibfield  {author} {\bibinfo {author} {\bibfnamefont {P.}~\bibnamefont {Dhara}}, \bibinfo {author} {\bibfnamefont {I.}~\bibnamefont {Padilla}}, \bibinfo {author} {\bibfnamefont {S.}~\bibnamefont {Guha}}, \bibinfo {author} {\bibfnamefont {A.}~\bibnamefont {Das}},\ and\ \bibinfo {author} {\bibfnamefont {K.}~\bibnamefont {Sinha}},\ }\bibfield  {title} {\bibinfo {title} {Non-{M}arkovian delay-assisted sensing with waveguide-coupled quantum emitters},\ }\href@noop {} {\bibfield  {journal} {\bibinfo  {journal} {arXiv:2605.05434}\ } (\bibinfo {year} {2026})}\BibitemShut {NoStop}%
\bibitem [{\citenamefont {Mirhosseini}\ \emph {et~al.}(2019)\citenamefont {Mirhosseini}, \citenamefont {Kim}, \citenamefont {Zhang}, \citenamefont {Sipahigil}, \citenamefont {Dieterle}, \citenamefont {Keller}, \citenamefont {Asenjo-Garcia}, \citenamefont {Chang},\ and\ \citenamefont {Painter}}]{Mirhosseini19}%
  \BibitemOpen
  \bibfield  {author} {\bibinfo {author} {\bibfnamefont {M.}~\bibnamefont {Mirhosseini}}, \bibinfo {author} {\bibfnamefont {E.}~\bibnamefont {Kim}}, \bibinfo {author} {\bibfnamefont {X.}~\bibnamefont {Zhang}}, \bibinfo {author} {\bibfnamefont {A.}~\bibnamefont {Sipahigil}}, \bibinfo {author} {\bibfnamefont {P.~B.}\ \bibnamefont {Dieterle}}, \bibinfo {author} {\bibfnamefont {A.~J.}\ \bibnamefont {Keller}}, \bibinfo {author} {\bibfnamefont {A.}~\bibnamefont {Asenjo-Garcia}}, \bibinfo {author} {\bibfnamefont {D.~E.}\ \bibnamefont {Chang}},\ and\ \bibinfo {author} {\bibfnamefont {O.}~\bibnamefont {Painter}},\ }\bibfield  {title} {\bibinfo {title} {Cavity quantum electrodynamics with atom-like mirrors},\ }\href {https://doi.org/10.1038/s41586-019-1196-1} {\bibfield  {journal} {\bibinfo  {journal} {Nature}\ }\textbf {\bibinfo {volume} {569}},\ \bibinfo {pages} {692} (\bibinfo {year} {2019})}\BibitemShut {NoStop}%
\bibitem [{\citenamefont {Keefe}\ \emph {et~al.}(2025)\citenamefont {Keefe}, \citenamefont {Agarwal},\ and\ \citenamefont {Kamal}}]{Keefe25}%
  \BibitemOpen
  \bibfield  {author} {\bibinfo {author} {\bibfnamefont {A.}~\bibnamefont {Keefe}}, \bibinfo {author} {\bibfnamefont {N.}~\bibnamefont {Agarwal}},\ and\ \bibinfo {author} {\bibfnamefont {A.}~\bibnamefont {Kamal}},\ }\bibfield  {title} {\bibinfo {title} {Quantifying spectral signatures of non-{M}arkovianity beyond the {B}orn-{R}edfield master equation},\ }\href {https://doi.org/10.22331/q-2025-09-24-1863} {\bibfield  {journal} {\bibinfo  {journal} {{Quantum}}\ }\textbf {\bibinfo {volume} {9}},\ \bibinfo {pages} {1863} (\bibinfo {year} {2025})}\BibitemShut {NoStop}%
\bibitem [{\citenamefont {Paulisch}\ \emph {et~al.}(2019)\citenamefont {Paulisch}, \citenamefont {Perarnau-Llobet}, \citenamefont {Gonz\'alez-Tudela},\ and\ \citenamefont {Cirac}}]{Paulisch19}%
  \BibitemOpen
  \bibfield  {author} {\bibinfo {author} {\bibfnamefont {V.}~\bibnamefont {Paulisch}}, \bibinfo {author} {\bibfnamefont {M.}~\bibnamefont {Perarnau-Llobet}}, \bibinfo {author} {\bibfnamefont {A.}~\bibnamefont {Gonz\'alez-Tudela}},\ and\ \bibinfo {author} {\bibfnamefont {J.~I.}\ \bibnamefont {Cirac}},\ }\bibfield  {title} {\bibinfo {title} {Quantum metrology with one-dimensional superradiant photonic states},\ }\href {https://doi.org/10.1103/PhysRevA.99.043807} {\bibfield  {journal} {\bibinfo  {journal} {Phys. Rev. A}\ }\textbf {\bibinfo {volume} {99}},\ \bibinfo {pages} {043807} (\bibinfo {year} {2019})}\BibitemShut {NoStop}%
\bibitem [{\citenamefont {Koppenh\"ofer}\ \emph {et~al.}(2022)\citenamefont {Koppenh\"ofer}, \citenamefont {Groszkowski}, \citenamefont {Lau},\ and\ \citenamefont {Clerk}}]{Koppenhofer22}%
  \BibitemOpen
  \bibfield  {author} {\bibinfo {author} {\bibfnamefont {M.}~\bibnamefont {Koppenh\"ofer}}, \bibinfo {author} {\bibfnamefont {P.}~\bibnamefont {Groszkowski}}, \bibinfo {author} {\bibfnamefont {H.-K.}\ \bibnamefont {Lau}},\ and\ \bibinfo {author} {\bibfnamefont {A.}~\bibnamefont {Clerk}},\ }\bibfield  {title} {\bibinfo {title} {Dissipative superradiant spin amplifier for enhanced quantum sensing},\ }\href {https://doi.org/10.1103/PRXQuantum.3.030330} {\bibfield  {journal} {\bibinfo  {journal} {PRX Quantum}\ }\textbf {\bibinfo {volume} {3}},\ \bibinfo {pages} {030330} (\bibinfo {year} {2022})}\BibitemShut {NoStop}%
\bibitem [{\citenamefont {Bohr}\ \emph {et~al.}(2024)\citenamefont {Bohr}, \citenamefont {Kristensen}, \citenamefont {Hotter}, \citenamefont {Sch{\"a}ffer}, \citenamefont {Robinson-Tait}, \citenamefont {Thomsen}, \citenamefont {Zelevinsky}, \citenamefont {Ritsch},\ and\ \citenamefont {M{\"u}ller}}]{Bohr24}%
  \BibitemOpen
  \bibfield  {author} {\bibinfo {author} {\bibfnamefont {E.~A.}\ \bibnamefont {Bohr}}, \bibinfo {author} {\bibfnamefont {S.~L.}\ \bibnamefont {Kristensen}}, \bibinfo {author} {\bibfnamefont {C.}~\bibnamefont {Hotter}}, \bibinfo {author} {\bibfnamefont {S.~A.}\ \bibnamefont {Sch{\"a}ffer}}, \bibinfo {author} {\bibfnamefont {J.}~\bibnamefont {Robinson-Tait}}, \bibinfo {author} {\bibfnamefont {J.~W.}\ \bibnamefont {Thomsen}}, \bibinfo {author} {\bibfnamefont {T.}~\bibnamefont {Zelevinsky}}, \bibinfo {author} {\bibfnamefont {H.}~\bibnamefont {Ritsch}},\ and\ \bibinfo {author} {\bibfnamefont {J.~H.}\ \bibnamefont {M{\"u}ller}},\ }\bibfield  {title} {\bibinfo {title} {Collectively enhanced {R}amsey readout by cavity sub- to superradiant transition},\ }\href {https://doi.org/10.1038/s41467-024-45420-x} {\bibfield  {journal} {\bibinfo  {journal} {Nat. Commun.}\ }\textbf {\bibinfo {volume} {15}},\ \bibinfo {pages} {1084} (\bibinfo {year} {2024})}\BibitemShut {NoStop}%
\bibitem [{\citenamefont {Castells-Graells}\ \emph {et~al.}(2025)\citenamefont {Castells-Graells}, \citenamefont {Cirac},\ and\ \citenamefont {Wild}}]{CastellsGraells25}%
  \BibitemOpen
  \bibfield  {author} {\bibinfo {author} {\bibfnamefont {D.}~\bibnamefont {Castells-Graells}}, \bibinfo {author} {\bibfnamefont {J.~I.}\ \bibnamefont {Cirac}},\ and\ \bibinfo {author} {\bibfnamefont {D.~S.}\ \bibnamefont {Wild}},\ }\bibfield  {title} {\bibinfo {title} {Cavity quantum electrodynamics with atom arrays in free space},\ }\href {https://doi.org/10.1103/PhysRevA.111.053712} {\bibfield  {journal} {\bibinfo  {journal} {Phys. Rev. A}\ }\textbf {\bibinfo {volume} {111}},\ \bibinfo {pages} {053712} (\bibinfo {year} {2025})}\BibitemShut {NoStop}%
\bibitem [{\citenamefont {Zafra-Bono}\ \emph {et~al.}(2026)\citenamefont {Zafra-Bono}, \citenamefont {Rubies-Bigorda},\ and\ \citenamefont {Yelin}}]{Zafra26}%
  \BibitemOpen
  \bibfield  {author} {\bibinfo {author} {\bibfnamefont {D.}~\bibnamefont {Zafra-Bono}}, \bibinfo {author} {\bibfnamefont {O.}~\bibnamefont {Rubies-Bigorda}},\ and\ \bibinfo {author} {\bibfnamefont {S.~F.}\ \bibnamefont {Yelin}},\ }\bibfield  {title} {\bibinfo {title} {Subradiant collective states for precision sensing via transmission spectra},\ }\href {https://doi.org/10.1103/rprq-bp61} {\bibfield  {journal} {\bibinfo  {journal} {Phys. Rev. A}\ }\textbf {\bibinfo {volume} {114}},\ \bibinfo {pages} {013704} (\bibinfo {year} {2026})}\BibitemShut {NoStop}%
\bibitem [{\citenamefont {Zanardi}\ and\ \citenamefont {Rasetti}(1997)}]{Zanardi97}%
  \BibitemOpen
  \bibfield  {author} {\bibinfo {author} {\bibfnamefont {P.}~\bibnamefont {Zanardi}}\ and\ \bibinfo {author} {\bibfnamefont {M.}~\bibnamefont {Rasetti}},\ }\bibfield  {title} {\bibinfo {title} {Noiseless quantum codes},\ }\href {https://doi.org/10.1103/PhysRevLett.79.3306} {\bibfield  {journal} {\bibinfo  {journal} {Phys. Rev. Lett.}\ }\textbf {\bibinfo {volume} {79}},\ \bibinfo {pages} {3306} (\bibinfo {year} {1997})}\BibitemShut {NoStop}%
\bibitem [{\citenamefont {Lidar}\ \emph {et~al.}(1998)\citenamefont {Lidar}, \citenamefont {Chuang},\ and\ \citenamefont {Whaley}}]{Lidar98}%
  \BibitemOpen
  \bibfield  {author} {\bibinfo {author} {\bibfnamefont {D.~A.}\ \bibnamefont {Lidar}}, \bibinfo {author} {\bibfnamefont {I.~L.}\ \bibnamefont {Chuang}},\ and\ \bibinfo {author} {\bibfnamefont {K.~B.}\ \bibnamefont {Whaley}},\ }\bibfield  {title} {\bibinfo {title} {Decoherence-free subspaces for quantum computation},\ }\href {https://doi.org/10.1103/PhysRevLett.81.2594} {\bibfield  {journal} {\bibinfo  {journal} {Phys. Rev. Lett.}\ }\textbf {\bibinfo {volume} {81}},\ \bibinfo {pages} {2594} (\bibinfo {year} {1998})}\BibitemShut {NoStop}%
\bibitem [{\citenamefont {Shi}\ \emph {et~al.}(2015)\citenamefont {Shi}, \citenamefont {Chang},\ and\ \citenamefont {Cirac}}]{Shi15}%
  \BibitemOpen
  \bibfield  {author} {\bibinfo {author} {\bibfnamefont {T.}~\bibnamefont {Shi}}, \bibinfo {author} {\bibfnamefont {D.~E.}\ \bibnamefont {Chang}},\ and\ \bibinfo {author} {\bibfnamefont {J.~I.}\ \bibnamefont {Cirac}},\ }\bibfield  {title} {\bibinfo {title} {Multiphoton-scattering theory and generalized master equations},\ }\href {https://doi.org/10.1103/PhysRevA.92.053834} {\bibfield  {journal} {\bibinfo  {journal} {Phys. Rev. A}\ }\textbf {\bibinfo {volume} {92}},\ \bibinfo {pages} {053834} (\bibinfo {year} {2015})}\BibitemShut {NoStop}%
\bibitem [{\citenamefont {Arranz~Regidor}\ \emph {et~al.}(2021)\citenamefont {Arranz~Regidor}, \citenamefont {Crowder}, \citenamefont {Carmichael},\ and\ \citenamefont {Hughes}}]{ArranzRegidor21}%
  \BibitemOpen
  \bibfield  {author} {\bibinfo {author} {\bibfnamefont {S.}~\bibnamefont {Arranz~Regidor}}, \bibinfo {author} {\bibfnamefont {G.}~\bibnamefont {Crowder}}, \bibinfo {author} {\bibfnamefont {H.}~\bibnamefont {Carmichael}},\ and\ \bibinfo {author} {\bibfnamefont {S.}~\bibnamefont {Hughes}},\ }\bibfield  {title} {\bibinfo {title} {Modeling quantum light-matter interactions in waveguide {QED} with retardation, nonlinear interactions, and a time-delayed feedback: Matrix product states versus a space-discretized waveguide model},\ }\href {https://doi.org/10.1103/PhysRevResearch.3.023030} {\bibfield  {journal} {\bibinfo  {journal} {Phys. Rev. Res.}\ }\textbf {\bibinfo {volume} {3}},\ \bibinfo {pages} {023030} (\bibinfo {year} {2021})}\BibitemShut {NoStop}%
\bibitem [{\citenamefont {Lanuza}\ and\ \citenamefont {Schneble}(2024)}]{Lanuza24}%
  \BibitemOpen
  \bibfield  {author} {\bibinfo {author} {\bibfnamefont {A.}~\bibnamefont {Lanuza}}\ and\ \bibinfo {author} {\bibfnamefont {D.}~\bibnamefont {Schneble}},\ }\bibfield  {title} {\bibinfo {title} {Exact solution for the collective non-{M}arkovian decay of two fully excited quantum emitters},\ }\href {https://doi.org/10.1103/PhysRevResearch.6.033196} {\bibfield  {journal} {\bibinfo  {journal} {Phys. Rev. Res.}\ }\textbf {\bibinfo {volume} {6}},\ \bibinfo {pages} {033196} (\bibinfo {year} {2024})}\BibitemShut {NoStop}%
\bibitem [{\citenamefont {Vodenkova}\ and\ \citenamefont {Pichler}(2024)}]{Vodenkova24}%
  \BibitemOpen
  \bibfield  {author} {\bibinfo {author} {\bibfnamefont {K.}~\bibnamefont {Vodenkova}}\ and\ \bibinfo {author} {\bibfnamefont {H.}~\bibnamefont {Pichler}},\ }\bibfield  {title} {\bibinfo {title} {Continuous coherent quantum feedback with time delays: Tensor network solution},\ }\href {https://doi.org/10.1103/PhysRevX.14.031043} {\bibfield  {journal} {\bibinfo  {journal} {Phys. Rev. X}\ }\textbf {\bibinfo {volume} {14}},\ \bibinfo {pages} {031043} (\bibinfo {year} {2024})}\BibitemShut {NoStop}%
\bibitem [{\citenamefont {Guimond}\ \emph {et~al.}(2016)\citenamefont {Guimond}, \citenamefont {Roulet}, \citenamefont {Le},\ and\ \citenamefont {Scarani}}]{Guimond16}%
  \BibitemOpen
  \bibfield  {author} {\bibinfo {author} {\bibfnamefont {P.-O.}\ \bibnamefont {Guimond}}, \bibinfo {author} {\bibfnamefont {A.}~\bibnamefont {Roulet}}, \bibinfo {author} {\bibfnamefont {H.~N.}\ \bibnamefont {Le}},\ and\ \bibinfo {author} {\bibfnamefont {V.}~\bibnamefont {Scarani}},\ }\bibfield  {title} {\bibinfo {title} {{R}abi oscillation in a quantum cavity: {M}arkovian and non-{M}arkovian dynamics},\ }\href {https://doi.org/10.1103/PhysRevA.93.023808} {\bibfield  {journal} {\bibinfo  {journal} {Phys. Rev. A}\ }\textbf {\bibinfo {volume} {93}},\ \bibinfo {pages} {023808} (\bibinfo {year} {2016})}\BibitemShut {NoStop}%
\bibitem [{\citenamefont {Burgess}\ and\ \citenamefont {Florescu}(2022)}]{Burgess22}%
  \BibitemOpen
  \bibfield  {author} {\bibinfo {author} {\bibfnamefont {A.}~\bibnamefont {Burgess}}\ and\ \bibinfo {author} {\bibfnamefont {M.}~\bibnamefont {Florescu}},\ }\bibfield  {title} {\bibinfo {title} {Non-{M}arkovian dynamics of a single excitation within many-body dissipative systems},\ }\href {https://doi.org/10.1103/PhysRevA.105.062207} {\bibfield  {journal} {\bibinfo  {journal} {Phys. Rev. A}\ }\textbf {\bibinfo {volume} {105}},\ \bibinfo {pages} {062207} (\bibinfo {year} {2022})}\BibitemShut {NoStop}%
\bibitem [{\citenamefont {Das}\ \emph {et~al.}(2025)\citenamefont {Das}, \citenamefont {Solano},\ and\ \citenamefont {Sinha}}]{Das25}%
  \BibitemOpen
  \bibfield  {author} {\bibinfo {author} {\bibfnamefont {A.}~\bibnamefont {Das}}, \bibinfo {author} {\bibfnamefont {P.}~\bibnamefont {Solano}},\ and\ \bibinfo {author} {\bibfnamefont {K.}~\bibnamefont {Sinha}},\ }\bibfield  {title} {\bibinfo {title} {Non-{M}arkovian spontaneous emission in a tunable cavity formed by atomic mirrors},\ }\href {https://doi.org/10.1103/29yv-12sq} {\bibfield  {journal} {\bibinfo  {journal} {Phys. Rev. A}\ }\textbf {\bibinfo {volume} {112}},\ \bibinfo {pages} {043723} (\bibinfo {year} {2025})}\BibitemShut {NoStop}%
\bibitem [{\citenamefont {Gonz{\'{a}}lez}\ and\ \citenamefont {Rivas}(2026)}]{Gonzalez26}%
  \BibitemOpen
  \bibfield  {author} {\bibinfo {author} {\bibfnamefont {I.}~\bibnamefont {Gonz{\'{a}}lez}}\ and\ \bibinfo {author} {\bibfnamefont {{\'{A}}.}~\bibnamefont {Rivas}},\ }\bibfield  {title} {\bibinfo {title} {From {S}uperradiance to {S}uperabsorption: {A}n {E}xact {T}reatment of {N}on-{M}arkovian {C}ooperative {R}adiation},\ }\href {https://doi.org/10.22331/q-2026-07-15-2159} {\bibfield  {journal} {\bibinfo  {journal} {{Quantum}}\ }\textbf {\bibinfo {volume} {10}},\ \bibinfo {pages} {2159} (\bibinfo {year} {2026})}\BibitemShut {NoStop}%
\bibitem [{\citenamefont {Rui}\ \emph {et~al.}(2020)\citenamefont {Rui}, \citenamefont {Wei}, \citenamefont {Rubio-Abadal}, \citenamefont {Hollerith}, \citenamefont {Zeiher}, \citenamefont {Stamper-Kurn}, \citenamefont {Gross},\ and\ \citenamefont {Bloch}}]{Rui20}%
  \BibitemOpen
  \bibfield  {author} {\bibinfo {author} {\bibfnamefont {J.}~\bibnamefont {Rui}}, \bibinfo {author} {\bibfnamefont {D.}~\bibnamefont {Wei}}, \bibinfo {author} {\bibfnamefont {A.}~\bibnamefont {Rubio-Abadal}}, \bibinfo {author} {\bibfnamefont {S.}~\bibnamefont {Hollerith}}, \bibinfo {author} {\bibfnamefont {J.}~\bibnamefont {Zeiher}}, \bibinfo {author} {\bibfnamefont {D.~M.}\ \bibnamefont {Stamper-Kurn}}, \bibinfo {author} {\bibfnamefont {C.}~\bibnamefont {Gross}},\ and\ \bibinfo {author} {\bibfnamefont {I.}~\bibnamefont {Bloch}},\ }\bibfield  {title} {\bibinfo {title} {A subradiant optical mirror formed by a single structured atomic layer},\ }\href {https://doi.org/10.1038/s41586-020-2463-x} {\bibfield  {journal} {\bibinfo  {journal} {Nature}\ }\textbf {\bibinfo {volume} {583}},\ \bibinfo {pages} {369} (\bibinfo {year} {2020})}\BibitemShut {NoStop}%
\bibitem [{\citenamefont {Guerin}\ \emph {et~al.}(2016)\citenamefont {Guerin}, \citenamefont {Ara\'ujo},\ and\ \citenamefont {Kaiser}}]{Guerin16}%
  \BibitemOpen
  \bibfield  {author} {\bibinfo {author} {\bibfnamefont {W.}~\bibnamefont {Guerin}}, \bibinfo {author} {\bibfnamefont {M.~O.}\ \bibnamefont {Ara\'ujo}},\ and\ \bibinfo {author} {\bibfnamefont {R.}~\bibnamefont {Kaiser}},\ }\bibfield  {title} {\bibinfo {title} {Subradiance in a large cloud of cold atoms},\ }\href {https://doi.org/10.1103/PhysRevLett.116.083601} {\bibfield  {journal} {\bibinfo  {journal} {Phys. Rev. Lett.}\ }\textbf {\bibinfo {volume} {116}},\ \bibinfo {pages} {083601} (\bibinfo {year} {2016})}\BibitemShut {NoStop}%
\bibitem [{\citenamefont {Ferioli}\ \emph {et~al.}(2021)\citenamefont {Ferioli}, \citenamefont {Glicenstein}, \citenamefont {Henriet}, \citenamefont {Ferrier-Barbut},\ and\ \citenamefont {Browaeys}}]{Ferioli21PRX}%
  \BibitemOpen
  \bibfield  {author} {\bibinfo {author} {\bibfnamefont {G.}~\bibnamefont {Ferioli}}, \bibinfo {author} {\bibfnamefont {A.}~\bibnamefont {Glicenstein}}, \bibinfo {author} {\bibfnamefont {L.}~\bibnamefont {Henriet}}, \bibinfo {author} {\bibfnamefont {I.}~\bibnamefont {Ferrier-Barbut}},\ and\ \bibinfo {author} {\bibfnamefont {A.}~\bibnamefont {Browaeys}},\ }\bibfield  {title} {\bibinfo {title} {Storage and release of subradiant excitations in a dense atomic cloud},\ }\href {https://doi.org/10.1103/PhysRevX.11.021031} {\bibfield  {journal} {\bibinfo  {journal} {Phys. Rev. X}\ }\textbf {\bibinfo {volume} {11}},\ \bibinfo {pages} {021031} (\bibinfo {year} {2021})}\BibitemShut {NoStop}%
\bibitem [{\citenamefont {Douglas}\ \emph {et~al.}(2026)\citenamefont {Douglas}, \citenamefont {Su}, \citenamefont {Szurek}, \citenamefont {Groth}, \citenamefont {Brandstetter}, \citenamefont {Markovi\'{c}}, \citenamefont {Rubies-Bigorda}, \citenamefont {Ostermann}, \citenamefont {Yelin},\ and\ \citenamefont {Greiner}}]{Douglas26arxiv}%
  \BibitemOpen
  \bibfield  {author} {\bibinfo {author} {\bibfnamefont {A.}~\bibnamefont {Douglas}}, \bibinfo {author} {\bibfnamefont {L.}~\bibnamefont {Su}}, \bibinfo {author} {\bibfnamefont {M.}~\bibnamefont {Szurek}}, \bibinfo {author} {\bibfnamefont {R.}~\bibnamefont {Groth}}, \bibinfo {author} {\bibfnamefont {S.}~\bibnamefont {Brandstetter}}, \bibinfo {author} {\bibfnamefont {O.}~\bibnamefont {Markovi\'{c}}}, \bibinfo {author} {\bibfnamefont {O.}~\bibnamefont {Rubies-Bigorda}}, \bibinfo {author} {\bibfnamefont {S.}~\bibnamefont {Ostermann}}, \bibinfo {author} {\bibfnamefont {S.~F.}\ \bibnamefont {Yelin}},\ and\ \bibinfo {author} {\bibfnamefont {M.}~\bibnamefont {Greiner}},\ }\bibfield  {title} {\bibinfo {title} {Many-body super- and subradiance in ordered atomic arrays},\ }\href@noop {} {\bibfield  {journal} {\bibinfo  {journal} {arXiv:2604.11795}\ } (\bibinfo {year} {2026})}\BibitemShut {NoStop}%
\bibitem [{\citenamefont {Mok}\ \emph {et~al.}(2024)\citenamefont {Mok}, \citenamefont {Poddar}, \citenamefont {Sierra}, \citenamefont {Rusconi}, \citenamefont {Preskill},\ and\ \citenamefont {Asenjo-Garcia}}]{Mok24arxiv}%
  \BibitemOpen
  \bibfield  {author} {\bibinfo {author} {\bibfnamefont {W.-K.}\ \bibnamefont {Mok}}, \bibinfo {author} {\bibfnamefont {A.}~\bibnamefont {Poddar}}, \bibinfo {author} {\bibfnamefont {E.}~\bibnamefont {Sierra}}, \bibinfo {author} {\bibfnamefont {C.~C.}\ \bibnamefont {Rusconi}}, \bibinfo {author} {\bibfnamefont {J.}~\bibnamefont {Preskill}},\ and\ \bibinfo {author} {\bibfnamefont {A.}~\bibnamefont {Asenjo-Garcia}},\ }\bibfield  {title} {\bibinfo {title} {Universal scaling laws for correlated decay of many-body quantum systems},\ }\href@noop {} {\bibfield  {journal} {\bibinfo  {journal} {arXiv:2406.00722}\ } (\bibinfo {year} {2024})}\BibitemShut {NoStop}%
\bibitem [{\citenamefont {Cardenas-Lopez}\ \emph {et~al.}(2023)\citenamefont {Cardenas-Lopez}, \citenamefont {Masson}, \citenamefont {Zager},\ and\ \citenamefont {Asenjo-Garcia}}]{CardenasLopez23}%
  \BibitemOpen
  \bibfield  {author} {\bibinfo {author} {\bibfnamefont {S.}~\bibnamefont {Cardenas-Lopez}}, \bibinfo {author} {\bibfnamefont {S.~J.}\ \bibnamefont {Masson}}, \bibinfo {author} {\bibfnamefont {Z.}~\bibnamefont {Zager}},\ and\ \bibinfo {author} {\bibfnamefont {A.}~\bibnamefont {Asenjo-Garcia}},\ }\bibfield  {title} {\bibinfo {title} {Many-body superradiance and dynamical mirror symmetry breaking in waveguide {QED}},\ }\href {https://doi.org/10.1103/PhysRevLett.131.033605} {\bibfield  {journal} {\bibinfo  {journal} {Phys. Rev. Lett.}\ }\textbf {\bibinfo {volume} {131}},\ \bibinfo {pages} {033605} (\bibinfo {year} {2023})}\BibitemShut {NoStop}%
\bibitem [{\citenamefont {Scully}\ and\ \citenamefont {Zubairy}(1997)}]{ScullyBook}%
  \BibitemOpen
  \bibfield  {author} {\bibinfo {author} {\bibfnamefont {M.~O.}\ \bibnamefont {Scully}}\ and\ \bibinfo {author} {\bibfnamefont {M.~S.}\ \bibnamefont {Zubairy}},\ }\href@noop {} {\emph {\bibinfo {title} {Quantum Optics}}}\ (\bibinfo  {publisher} {Cambridge University Press},\ \bibinfo {year} {1997})\BibitemShut {NoStop}%
\end{thebibliography}%

\qquad

\newpage

\onecolumngrid
\begin{center}
\large{\textbf{SUPPLEMENTAL MATERIAL}}
\end{center}

\section{Collective-delay organization}

We present here the analytical basis for the emergence of the collective memory parameter $\chi$ presented in the main text. First, we show the parameter organizes dynamics for certain states in the large $N$ limit; second, we derive the quasi-Markovian correction to the bright decay rate. We explicitly account for intrinsic losses and chiral decay. We take $\Gamma$ to be the total single-emitter decay rate and denote by $\beta_{L(R)}$ the waveguide coupling efficiency to modes propagating to the left (right). The guided and nonguided decay rates are consequently given by $\beta_{L(R)}\,\Gamma$ and $(1-\beta_L - \beta_R)\,\Gamma$, respectively. The ideal waveguide limit is recovered for $\beta_{L}+\beta_{R} = \beta = 1$, and a bidirectional waveguide is defined by a chirality $\zeta \equiv \beta_L / \beta_R = 1$. We formulate the analysis in dimensionless units of $T=\Gamma t$ and the single-excitation amplitude vector
\begin{equation}
    \bm C(T)=(c_1(T),c_2(T),\ldots,c_N(T))^{\top}.
\end{equation}
The dimensionless nearest-neighbor delay is $\eta=\,\Gamma\ell/(N-1)v_g$. Then the coupled-amplitude equations can be cast in matrix form as
\begin{align}
    \frac{d \bm C(T)}{dT}
    &=
    -\frac12\bm C(T)
    -
    \sum_{r=1}^{N-1}
    \mathbb S_r\,
    \bm C(T-r\eta)\,
    \Theta(T-r\eta),
    \label{eq:EM_matrix_delay_T}
\end{align}
where the $N\times N$ separation matrix $\mathbb S_r$ is defined by elements $(\mathbb S_r)_{mn}= \beta_L \delta_{|m-n|,r}$ for $n > m$ and $(\mathbb S_r)_{mn}= \beta_{R}\delta_{|m-n|,r}$ for $m>n$. This generalizes Eq.~\eqref{eq:delay_eom_inphase} for finite internal losses and chiral couplings. 

\subsection{Collective-delay organization for $N\gg 1$}
We explain why the collective delay $\chi=N\Gamma\ell/v_g$ organizes the dynamics for any family of states with a macroscopic continuum profile. Writing Eq.~\eqref{eq:EM_matrix_delay_T} in component form gives
\begin{equation}
    \frac{\mathrm{d}c_m(T)}{\mathrm{d}T}
=-\frac12 c_m(T)
 -\beta_R
\sum_{n < m}
c_n\big(T-\eta_{mn}\big)
\Theta\big(T-\eta_{mn})\big)- \beta_L
\sum_{n > m}
c_n\big(T-\eta_{mn}\big)
\Theta\big(T-\eta_{mn})\big),
\end{equation}
where $\eta_{mn}=\eta\,|m-n|$. Introducing the fractional coordinate  $\xi_m=(x_m-x_1)/\ell=(m-1)/(N-1)$, a collective time variable $\vartheta=\beta N T$ and the rescaled amplitudes ${A}_N(\xi_m,\vartheta) = \sqrt{N} c_{m}(T)$, the above equation transforms into
\begin{align}
\frac{\partial{A}_N(\xi_m,\vartheta)}
{\partial\vartheta}
=
-\frac{1}{2\beta N}
{A}_N(\xi_m,&\vartheta)
-\frac{1}{N(1+\zeta)}
\sum_{n< m} {A}_N\!\big(
\xi_n,
\vartheta-\beta\chi(\xi_m-\xi_n)
\big)
\Theta\big(
\vartheta-\beta\chi(\xi_m-\xi_n)
\big)\notag \\&~~~~~~~~~~~~~~~~~~~~~~~-\frac{\zeta}{N(1+\zeta)}
\sum_{n> m} {A}_N\!\big(
\xi_n,
\vartheta-\beta\chi(\xi_n-\xi_m)
\big)
\Theta\big(
\vartheta-\beta\chi(\xi_n-\xi_m)
\big).
\label{eq:finiteN_continuum}
\end{align}
In the limit $N\rightarrow\infty$, the first term is subleading and therefore can be dropped, while the finite sum can be promoted to a Riemann integral over the variable $\xi'\in(0,1)$. Then, for any family of states admitting a continuum limit, i.e., 
\begin{equation}
    {A}_N(\xi_m,\vartheta)\xrightarrow{N\rightarrow\infty,\xi_m\rightarrow\xi}{A}(\xi,\vartheta),
\end{equation}
the equation reduces to the instructive form
\begin{equation}
\frac{\partial{A}(\xi,\vartheta)}
{\partial\vartheta}\simeq-\frac{1}{1+\zeta}
\int_{0}^{\xi}\dd\xi'
{A}\!\big(
\xi',
\vartheta-\beta\chi(\xi-\xi')
\big)
\Theta\!\big(
\vartheta-\beta\chi(\xi-\xi')
\big) -\frac{\zeta}{1+\zeta}
\int_{\xi}^{1}\dd\xi'
{A}\!\big(
\xi',
\vartheta-\beta\chi(\xi'-\xi)
\big)
\Theta\!\big(
\vartheta-\beta\chi(\xi'-\xi)
\big).
\label{eq:finiteN_continuum}
\end{equation}
Eq.~\eqref{eq:finiteN_continuum} highlights how finite-delay modifications to the dynamics enter solely through the combination $\beta\chi$ and the chirality $\zeta$, without dependence on $N$ or $\ell$ separately. The appearance of $\beta N$ provides the optical depth as an effective emitter number for imperfect waveguide couplings. As long as the initial emitter amplitudes can be asymptotically mapped onto a well-defined density function ${A}_0(\xi)$, the ensuing dynamics take the form 
\begin{equation}
{A}_N(\xi_m,\vartheta)\xrightarrow{N\rightarrow\infty,\xi_m\rightarrow\xi}{A}_{{\beta\chi},\zeta}(\xi,\vartheta)\big|_{{A_0}},
\end{equation}
with the precise trajectory set by its initial profile $A_0(\xi)$.

The universality of behavior controlled solely by $\chi$ holds in the large-$N$ limit for a wide variety of states. The most obvious example is the superradiant state for which $A_0(\xi) = 1$, but the same condition holds for many subradiant states as well. For example, the double-domain structure considered in the main text is well-defined by $A_0(\xi) = 1 - 2\Theta(\xi-1/2)$. Because of this, such a state must have behavior scaling in the same manner as the superradiant state: non-Markovian signatures grow stronger with increasing $N$ and fixed $\ell$. On the contrary, the alternating sign structure considered in the main text cannot be well-defined in the asymptotic limit and hence is able to scale differently. Only states without well-defined asymptotic density functions can have delay-induced leakage errors with favorable scaling with system size.

\subsection{Weak-delay perturbation theory}

Here we derive the quasi-Markovian result in Eq.~\eqref{eq:SR_second_order_fixed_length}. We take the Laplace transform with respect to dimensionless time $T$ defining
\begin{equation}
    \widetilde{\bm{C}}(S)
    \equiv
    \mathcal L_T[\bm C(T)]
    =
    \int_0^\infty dT\,e^{-S T}\bm C(T),
\end{equation}
where $S$ is the Laplace variable conjugate to $T$.

Applying this transform to the dimensionless delay equation Eq.~\eqref{eq:EM_matrix_delay_T} gives (assuming no chirality, $\zeta=1$)
\begin{align}
    \widetilde{\bm C}(S)
    =
    \Bigg[
    \bigg(S+\frac12\bigg)\mathds{1}_N
    +
    \frac{\beta}{2}
    \bigg(\mathbb K_N(\rho)-\mathds{1}_N\bigg)
    \Bigg]^{-1}
    \bm C(0),
    \label{eq:EM_laplace_solution}
\end{align}
where $\rho=e^{-\eta S}$, $\mathds{1}_N$ denotes the $N\times N$ identity matrix, and $\mathbb K_N(\rho)$ represents a matrix with elements $\big(\mathbb K_N(\rho)\big)_{mn}=\rho^{|m-n|}$. In obtaining Eq.~\eqref{eq:EM_laplace_solution}, we assume no excitation history prior to $T=0$ and invoke the identity
\begin{equation}
    \mathcal L_T
    \big[
    c_n(T-r\eta)\Theta(T-r\eta)
    \big]
    =
    e^{-r\eta S}\,\widetilde c_n(S),
\end{equation}
where $r = |m-n|$.

For each eigenvalue branch $\lambda_j(\rho)$ of $\mathbb K_N(\rho)$, the corresponding pole satisfies
\begin{equation}
   S+\frac12
    +
    \frac{\beta}{2}
    \left[\lambda_j(\rho)-1\right]
    =
    0.
    \label{eq:EM_scalar_pole_condition}
\end{equation}
The bright branch is the pole continuously connected to the Markovian bright state $\ket B=N^{-1/2}\sum_m\ket {e_m}$. At zero delay, since $\lambda_B(1)=N$, the Markovian bright pole is given by $S_0=-[1+\beta(N-1)]/2$. Because this eigenvalue is nondegenerate, the corresponding eigenvalue branch can be expanded analytically for sufficiently small $\eta$. Consequently, an expansion of the associated pole can be expressed as
\begin{equation}
    S(\eta)
    =
    S_0
    +
    \eta S_1
    +
    \eta^2S_2
    +
    \mathcal O(\eta^3).
    \label{eq:EM_s_series}
\end{equation}
Since $\rho=e^{-\eta S(\eta)}$, its weak-delay expansion goes as
\begin{align}
    \rho
    =\sum_{r=0}^{\infty}\frac{(-\eta)^r(S(\eta))^r}{r!}=\sum_{r=0}^2\eta^rx_r+
    \mathcal O(\eta^3),
    \label{eq:EM_rho_series}
\end{align}
with $x_1=-S_0$ and $x_2=-S_1+S_0^2/2$. Finally, expanding the bright eigenvalue branch around $\rho=1$ gives
\begin{align}
    &\lambda_B(\rho)
    =\sum_{r=0}^{\infty}\frac{\lambda_B^{(r)}(1)}{r!}(\rho-1)^{r}\notag\\ 
    = &\; N
    +
    \eta\lambda'_B(1)x_1
    +
    \eta^2
    \bigg[
    \lambda'_B(1)x_2
    +
    \frac12\lambda''_B(1)x_1^2
    \bigg]
    +
    \mathcal O(\eta^3).
    \label{eq:EM_lambda_eta_expand}
\end{align}
Substituting Eqs.~\eqref{eq:EM_s_series} and \eqref{eq:EM_lambda_eta_expand} into the pole condition \eqref{eq:EM_scalar_pole_condition}, and matching powers of $\eta$, yields
\begin{subequations}
\begin{align}
    S_1
    &=
    \frac{\beta}{2}\lambda'_B(1)S_0,
    \label{eq:EM_s1_result} \\
    S_2
    &=
    \frac{\beta^2}{4}
    \big[\lambda'_B(1)\big]^2S_0
    -
    \frac{\beta}{4}
    \big[
    \lambda'_B(1)
    +
    \lambda''_B(1)
    \big]S_0^2 .
    \label{eq:EM_s2_result}
\end{align}
\end{subequations}
To evaluate the two eigenvalue derivatives entering this expression, we define $\mathbb K_N^{\prime}
    =
    \left.
    \partial_\rho\mathbb K_N
    \right|_{\rho=1}$ and $
    \mathbb K_N^{\prime\prime}
    =
    \left.
    \partial_{\rho^2}\mathbb K_N
    \right|_{\rho=1}$. 
Since $(\mathbb K_N)_{mn}=\rho^{|m-n|}$, the matrix elements are $(\mathbb K_N^{\prime})_{mn}=|m-n|$ and $(\mathbb K_N^{\prime\prime})_{mn}=|m-n|(|m-n|-1)$. Because $\mathbb K_N(\rho)$ is an analytic symmetric matrix family near $\rho=1$ and $\lambda_B(1)=N$ is a simple eigenvalue, standard eigenvalue perturbation theory yields:
\begin{align}
    \lambda'_B(1)
    &=
    \bm B^{\top}\mathbb K_N^{\prime}\bm B,
    \label{eq:EM_lambda_prime_RQ}
    \\
    \lambda''_B(1)
    &=
    \bm B^{\top}\mathbb K_N^{\prime\prime}\bm B
    +
    \frac{2}{N}\sum_{\mu\in{\rm dark}}
    \bigg|\bm D_\mu^{\top}\mathbb K_N^{\prime}\bm B\bigg|^2
    ,
    \label{eq:EM_lambda_second_fundamental}
\end{align}
where $\bm B=(1,1,\ldots,1)^{\top}/\sqrt{N}$ is the $N$-component amplitude vector corresponding to the bright state $\ket{B}$, and $\{\bm D_\mu\}$ represents any orthonormal basis of the Markovian dark manifold. We use $\sum_{\mu\in{\rm dark}}\bm D_\mu\bm D_\mu^{\top}=\mathds{1}_N-\bm B\bm B^{\top}$ to reduce Eq.~\eqref{eq:EM_lambda_second_fundamental} to
\begin{equation}
    \lambda''_B(1)
    =
    \bm B^{\top}\mathbb K_N^{\prime\prime}\bm B
    +
    \frac{2}{N}\bigg[
    \bm B^{\top}(\mathbb K_N^{\prime})^2\bm B
    -
    \big(\bm B^{\top}\mathbb K_N^{\prime}\bm B\big)^2
    \bigg].
    \label{eq:EM_lambda_second_projector}
\end{equation}

The first derivative is obtained by summation over all separations $r=|m-n|$. Since there are $2(N-r)$ ordered emitter pairs at separation $r$, we find
\begin{align}
    \lambda'_B(1)
    &=
    \frac{1}{N}\sum_{m,n=1}^N |m-n|
    =
    \frac{2}{N}\sum_{r=1}^{N-1} r(N-r)=
    \frac{N^2-1}{3}.
    \label{eq:EM_lambda_prime_exact}
\end{align}
Likewise, the direct term $\bm B^{\top}\mathbb K_N^{\prime\prime}\bm B$ and the projector term $\bm B^{\top}(\mathbb K_N^{\prime})^2\bm B-(\bm B^{\top}\mathbb K_N^{\prime}\bm B)^2$ are evaluated by the same finite-distance sums. This yields
\begin{align}
    \lambda''_B(1)
    &=
    \frac{(N-2)(N-1)(N+1)(8N+1)}
         {45N}.
    \label{eq:EM_lambda_second_exact}
\end{align}

For the real bright pole considered here, the population decay rate is obtained as $\Gamma_B(\eta)=-2\Gamma\,S(\eta)$. Substituting Eqs.~\eqref{eq:EM_lambda_prime_exact} and \eqref{eq:EM_lambda_second_exact} into Eqs.~\eqref{eq:EM_s1_result} and \eqref{eq:EM_s2_result} gives the corresponding asymptotic expansion $(N\gg 1)$
\begin{align}
    \frac{\Gamma_B(\eta)}{N\beta\Gamma}
    =1+\frac{\beta\chi}{6}+\frac{(\beta\chi)^2}{20}+\mathcal{O}((\beta\chi)^3).
    \label{eq:EM_eta1_rate_lossy}
\end{align}

\section{Dynamics with arbitrary emitter spacing}

\begin{figure*}
    \includegraphics[width=\textwidth]{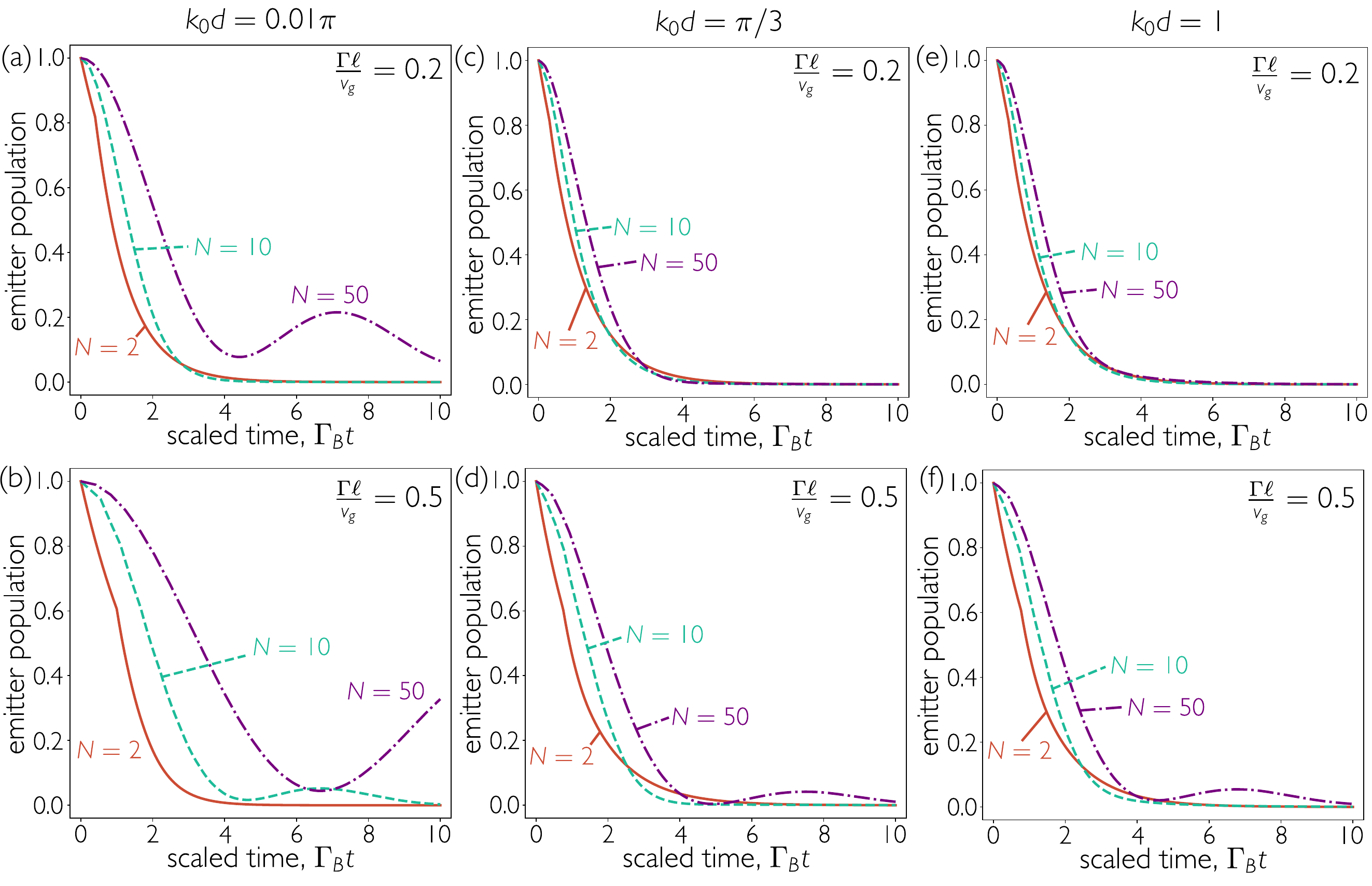}
    \caption{Emergent non-Markovianity away from the phase-matched condition. Time evolution of total emitter population, $\sum_{m=1}^N |c_m(t)|^2$, for different emitter numbers. In the top row $\Gamma\ell / v_g = 0.2$ while in the bottom row $\Gamma\ell / v_g = 0.5$. In (a,b) $k_0d = 0.01\pi$, (c,d) $k_0d = \pi/3$, and in (e,f) $k_0d = 1$. All plots use an initial state corresponding to the Markovian single-excitation superradiant state and scale time via that superradiant state's lifetime.}
    \label{fig: notbragg}
\end{figure*}

In the main text, we assume that the interemitter spacing always satisfies the phase-matched condition: $k_0d=2\pi q$. Relaxing this phase-matched array condition means that the components of the single-excitation amplitude vector now evolve according to
\begin{equation}
    \frac{\mathrm{d}c_m(T)}{\mathrm{d}T}
=-\frac12 c_m(T)
 -\beta_R
\sum_{n < m} \mathrm{e}^{\ii k_0 (x_m - x_n)}
c_n\big(T-\eta_{mn}\big)
\Theta\big(T-\eta_{mn})\big)- \beta_L
\sum_{n > m} \mathrm{e}^{\ii k_0 (x_n - x_m)}
c_n\big(T-\eta_{mn}\big)
\Theta\big(T-\eta_{mn})\big).
\end{equation}
Following the same steps as for the derivation for phase-matched arrays, we arrive at
\begin{align}
\frac{\partial{A}_N(\xi_m,\vartheta)}
{\partial\vartheta}
=
-\frac{1}{2\beta N}
{A}_N(\xi_m,\vartheta)
-&\frac{1}{N(1+\zeta)}
\sum_{n< m} \mathrm{e}^{\ii k_0 \ell (\xi_m - \xi_n)} {A}_N\!\big(
\xi_n,
\vartheta-\beta\chi(\xi_m-\xi_n)
\big)
\Theta\big(
\vartheta-\beta\chi(\xi_m-\xi_n)
\big)\notag \\-&\frac{\zeta}{N(1+\zeta)}
\sum_{n> m} \mathrm{e}^{\ii k_0\ell (\xi_n - \xi_m)}{A}_N\!\big(
\xi_n,
\vartheta-\beta\chi(\xi_n-\xi_m)
\big)
\Theta\big(
\vartheta-\beta\chi(\xi_n-\xi_m)
\big).
\label{eq:finiteN_continuum}
\end{align}
In the limit $N\rightarrow\infty$, we can again drop the first term and promote the finite sums to Riemann integrals over the variable $\xi'\in(0,1)$ giving
\begin{align}
\frac{\partial{A}_N(\xi,\vartheta)}
{\partial\vartheta}\simeq-\frac{1}{1+\zeta}
\int_{0}^{\xi}\dd\xi' \mathrm{e}^{\ii k_0 \ell (\xi - \xi')}
{A}_N\!\big(
\xi',
&\vartheta-\beta\chi(\xi-\xi')
\big)
\Theta\!\big(
\vartheta-\beta\chi(\xi-\xi')
\big)\notag \\&-\frac{\zeta}{1+\zeta}
\int_{\xi}^{1}\dd\xi' \mathrm{e}^{\ii k_0 \ell (\xi' - \xi)}
{A}_N\!\big(
\xi',
\vartheta-\beta\chi(\xi'-\xi)
\big)
\Theta\!\big(
\vartheta-\beta\chi(\xi'-\xi)
\big).
\label{eq:finiteN_continuum_notbragg}
\end{align}
The dynamics depends on both $\ell$ and $\chi$ independently, and is not universally described by the collective lifetime ratio and the chirality. One exception is for perfect chirality: if $\zeta = 0$ (or equivalently $\zeta \rightarrow\infty$) then only the second integral remains and a phase-shifted fractional distance $\xi \rightarrow \mathrm{e}^{\ii k_0 \ell \xi}\xi$ can be defined to absorb all $\ell$ dependence into the amplitudes. Note that in this case the population dynamics are equivalent since they depend on $|c_m(t)|^2$, but that does not necessarily mean the scattered fields are equivalent as those depend on the amplitudes directly.

Non-Markovianity is an emergent property regardless of the emitter spacing. As shown in Fig.~\ref{fig: notbragg}, at all emitter spacings initially Markovian behavior is eventually replaced by memory-altered dynamics as $N$ increases. At small spacings close to the phase-matched condition, non-Markovianity emerges at smaller emitter numbers than at larger spacings. Likely this is because the superradiant lifetime is approximately halved away from the phase-matched condition due to the presence of two competing loss channels~\cite{CardenasLopez23}, lowering the relevant system timescale. The role of $\beta\chi$ in Eq.~\eqref{eq:finiteN_continuum_notbragg} suggests that the relevant parameters governing this emergent non-Markovianity are, as in the phase-matched condition, the superradiant lifetime and the end-to-end delay length.

\section{Numerical results for constant $\beta\chi$ with imperfect waveguide coupling}

\begin{figure*}
    \includegraphics[width=\textwidth]{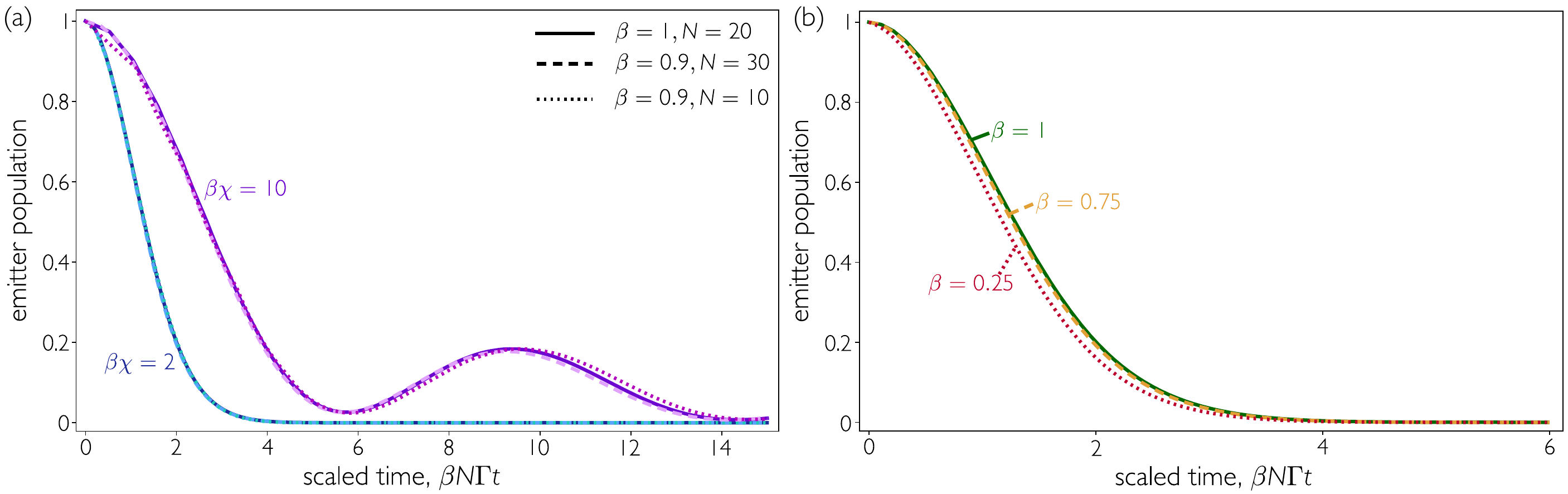}
    \caption{Dependence on $\beta\chi$. (a) Time evolution of total emitter population, $\sum_{m=1}^N |c_m(t)|^2$, for different values of $\beta\chi = \beta N\Gamma\ell/v_g$. (b) Time evolution of total emitter population for $\beta\chi = 2$ with $\beta=\set{1,0.75,0.25}$ and $N=\set{20,30,40}$. In both plots, the emitters form a phase-matched array and are initialized in the single-excitation superradiant state $\ket{\psi(t=0)} = \hat{S}_+ \ket{g}^{\otimes N} / \sqrt{N}$.}
    \label{fig: changebeta}
\end{figure*}

Relaxing the assumption of perfect coupling to the waveguide generalizes the dependence on $\chi$ in the main text to a dependence on $\beta\chi$, where $\beta$ is the waveguide coupling efficiency. In Fig.~\ref{fig: changebeta}, we show that the behavior is well-captured by the product $\beta\chi$ for various values of $\beta \neq 1$. The result is valid for all values of $\beta$ but small $\beta$ has more significant finite-size effects due to an inflation of the importance of the subleading first term, i.e., self-decay.

\section{Intensity and power spectrum}

The right-going intensity is defined as the normally ordered field intensity 
\begin{align}
I_R(x,t)=\frac{\varepsilon_0 v_g}{2}\expval{\hat E_R^{(-)}(x,t)\hat E_R^{(+)}(x,t)},
\end{align}
with the positive-frequency right-moving field operator defined as $\hat E_R^{(+)}(x,t)=\int_0^\infty dk\,E_k\,\hat a(k)e^{ikx-i\omega(k)t}$, where $\hat E_R^{(-)}=[\hat E_R^{(+)}]^\dagger$. The averaging is performed over the time-dependent field-emitter state
    \begin{equation}
\ket{\Psi(t)}
=
\sum_{m=1}^{N} c_m(t)\,\hat{\sigma}_+^{(m)}\ket{g}^{\otimes N}\ket{0}
+
\int_0^\infty dk\,\left[
c_R(k,t)\ket{g}^{\otimes N}\ket{1_{R,k}}
+
c_L(k,t)\ket{g}^{\otimes N}\ket{1_{L,k}}
\right],
\end{equation}
with $R (L)$ signifies the right-(left-) moving field.

In the single-excitation sector under phase-matched conditions, assuming $E_k\approx E_{k_0}$ and linear dispersion, the intensity evaluated at the right edge of the array, $x=x_N$, reduces to a coherent sum of retarded emitter amplitudes~\cite{Sinha20},
\begin{equation}
I(t)
\propto
\left|
\sum_{m=1}^{N}
c_m\bigg(t-\frac{x_N-x_m}{v_g}\bigg)\,
\Theta\bigg(t-\frac{x_N-x_m}{v_g}\bigg)
\right|^2.
\end{equation}

The finite-time power spectrum of the emitted field at $x=x_N$ is defined as~\cite{ScullyBook}
\begin{equation}
S(\omega;t)
=
\frac{1}{2\pi}
\int_0^t dt_1\int_0^t dt_2\,
e^{i\omega(t_1-t_2)}
G^{(1)}(t_1,t_2),
\end{equation}
where $G^{(1)}(t_1,t_2)=\expval{\hat E^{(-)}(x_N,t_1)\hat E^{(+)}(x_N,t_2)}$ denotes the first-order coherence function. In the single-excitation sector and integrating over the entire emitted pulse duration, for the phase-matched configuration, this simplifies to
\begin{equation}
S(\omega)
=
\frac{|E_{k_0}|^2}{2\pi}
\lim_{t\rightarrow\infty}\left|
 \int_0^t dt'\,
\mathcal E(t')e^{i\omega t'}
\right|^2,
\end{equation}
where
\begin{align}
    \mathcal E(t)\propto
\sum_m c_m\bigg(t-\frac{x_N-x_m}{v_g}\bigg)\Theta\bigg(t-\frac{x_N-x_m}{v_g}\bigg).
\end{align}

\end{document}